\documentclass[twocolumn]{aastex63}

\usepackage{color,graphicx,natbib}
\usepackage{graphicx}
\usepackage{amsmath}
\usepackage{amssymb}
\usepackage{bm}
\usepackage[toc,page]{appendix}
\usepackage{xspace}

\def\ra#1#2#3{#1$^{\rm h}$#2$^{\rm m}$#3$^{\rm s}$}
\def\dec#1#2#3{$#1^\circ #2' #3''$}

\def\swift{{\it Swift}}

\def\grb{GRB\,231117A\xspace}

\shortauthors{Schroeder et al.}
\shorttitle{GRB 231117A and the Radio-Detected Short GRB Population}

\begin{document}

\author[0000-0001-9915-8147]{Genevieve~Schroeder}
\affiliation{Center for Interdisciplinary Exploration and Research in Astrophysics (CIERA) and Department of Physics and Astronomy, Northwestern University, Evanston, IL 60208, USA}

\author[0000-0002-7374-935X]{Wen-fai Fong}\affiliation{Center for Interdisciplinary Exploration and Research in Astrophysics (CIERA) and Department of Physics and Astronomy, Northwestern University, Evanston, IL 60208, USA}

\author[0000-0002-5740-7747]{Charles D. Kilpatrick}\affiliation{Center for Interdisciplinary Exploration and Research in Astrophysics (CIERA) and Department of Physics and Astronomy, Northwestern University, Evanston, IL 60208, USA}

\author[0000-0003-3937-0618]{Alicia Rouco Escorial}\affiliation{European Space Agency (ESA), European Space Astronomy Centre (ESAC), Camino Bajo del Castillo s/n, 28692 Villanueva de la Cañada, Madrid, Spain}

\author[0000-0003-1792-2338]{Tanmoy Laskar}
\affiliation{Department of Physics \& Astronomy, University of Utah, Salt Lake City, UT 84112, USA}

\author[0000-0002-2028-9329]{Anya E. Nugent}
\affiliation{Center for Interdisciplinary Exploration and Research in Astrophysics (CIERA) and Department of Physics and Astronomy, Northwestern University, Evanston, IL 60208, USA}

\author[0000-0002-9267-6213]{Jillian Rastinejad}\affiliation{Center for Interdisciplinary Exploration and Research in Astrophysics (CIERA) and Department of Physics and Astronomy, Northwestern University, Evanston, IL 60208, USA}

%%%%%%
%%%%%% Alphabetical
%%%%%%

\author[0000-0002-8297-2473]{Kate D.~Alexander}
\affiliation{Steward Observatory, University of Arizona, 933 North Cherry Avenue, Tucson, AZ 85721-0065, USA}

\author[0000-0002-9392-9681]{Edo Berger}
\affiliation{Center for Astrophysics | Harvard \& Smithsonian, Cambridge, MA 02138, USA}

\author[0000-0001-5955-2502]{Thomas~G.~Brink}
\affiliation{Department of Astronomy, University of California, Berkeley, CA 94720-3411, USA}

\author[0000-0002-7706-5668]{Ryan~Chornock}
\affiliation{Department of Astronomy, University of California, Berkeley, CA 94720-3411, USA}

\author{Clecio R. de Bom}
\affiliation{Centro Brasileiro de Pesquisas F\'{ı}sicas, Rua Dr. Xavier Sigaud 150, CEP 22290-180, Rio de Janeiro, RJ, Brazil}

\author[0000-0002-9363-8606]{Yuxin Dong}
\affiliation{Center for Interdisciplinary Exploration and Research in Astrophysics (CIERA) and Department of Physics and Astronomy, Northwestern University, Evanston, IL 60208, USA}

\author[0000-0003-0307-9984]{Tarraneh Eftekhari}
\altaffiliation{NHFP Einstein Fellow}
\affiliation{Center for Interdisciplinary Exploration and Research in Astrophysics (CIERA) and Department of Physics and Astronomy, Northwestern University, Evanston, IL 60208, USA}

\author[0000-0003-3460-0103]{Alexei V. Filippenko}
\affiliation{Department of Astronomy, University of California, Berkeley, CA 94720-3411, USA}

\author[0009-0008-7505-5707]{Celeste Fuentes-Carvajal}
\affiliation{Departamento de Astronom\'ia y Astrof\'isica, Pontificia Universidad Catol\'ica de Chile, Av. Vicuña Mackenna 4860, 782-0436 Macul, Santiago, Chile}

\author[0000-0002-3934-2644]{Wynn~V.~Jacobson-Gal\'{a}n}
\affiliation{Department of Astronomy, University of California, Berkeley, CA 94720-3411, USA}

\author[0000-0001-6919-1237]{Matthew Malkan}
\affiliation{Department of Physics and Astronomy, UCLA, Los Angeles, CA 90095-1547, USA}

\author[0000-0003-4768-7586]{Raffaella Margutti}
\affiliation{Department of Astronomy, University of California, Berkeley, CA 94720-3411, USA}
\affiliation{Department of Physics, University of California, 366 Physics North MC 7300,
Berkeley, CA 94720, USA}

 \author[0000-0002-0744-0047]{Jeniveve Pearson}
\affiliation{Steward Observatory, University of Arizona, 933 North Cherry Avenue, Tucson, AZ 85721-0065, USA}

\author[0000-0003-2705-4941]{Lauren~Rhodes}
\affiliation{Astrophysics, Department of Physics, University of Oxford, Keble Road, Oxford, OX1 3RH, UK}

\author[0000-0002-1206-1930]{Ricardo Salinas}
\affiliation{Departamento de Astronom\'ia, Universidad de La Serena, Av. Juan Cisternas 1200, La Serena, Chile}

\author[0000-0003-4102-380X]{David J.\ Sand}
\affiliation{Steward Observatory, University of Arizona, 933 North Cherry Avenue, Tucson, AZ 85721-0065, USA}

\author[0000-0003-3402-6164]{Luidhy Santana-Silva}
\affiliation{Centro Brasileiro de Pesquisas F\'{ı}sicas, Rua Dr. Xavier Sigaud 150, CEP 22290-180, Rio de
Janeiro, RJ, Brazil}

\author{Andre Santos}
\affiliation{Centro Brasileiro de Pesquisas F\'{ı}sicas, Rua Dr. Xavier Sigaud 150, CEP 22290-180, Rio de Janeiro, RJ, Brazil}

\author[0000-0001-8023-4912]{Huei Sears}\affiliation{Center for Interdisciplinary Exploration and Research in Astrophysics (CIERA) and Department of Physics and Astronomy, Northwestern University, Evanston, IL 60208, USA}

\author[0000-0002-4022-1874]{Manisha Shrestha}
\affiliation{Steward Observatory, University of Arizona, 933 North Cherry Avenue, Tucson, AZ 85721-0065, USA}

\author[0000-0001-5510-2424]{Nathan Smith}
\affiliation{Steward Observatory, University of Arizona, 933 North Cherry Avenue, Tucson, AZ 85721-0065, USA}

\author{Wayne Webb}
\affiliation{UCLA Department of Physics and Astronomy, Los Angeles, CA 90095-1547, USA}

\author[0000-0003-2449-1329]{Simon de Wet}
\affiliation{University of Cape Town Astronomy Department, Private Bag X3, Rondebosch, 7701, South Africa}

\author[0000-0002-6535-8500]{Yi Yang}
\affiliation{Physics Department, Tsinghua University, Beijing, 100084, China}
\affiliation{Department of Astronomy, University of California, Berkeley, CA 94720-3411, USA}

\title{The Long-lived Broadband Afterglow of Short Gamma-Ray Burst 231117A and the Growing Radio-Detected Short GRB Population}

\begin{abstract}
    We present multiwavelength observations of the {\it Swift} short $\gamma$-ray burst \grb, localized to an underlying galaxy at redshift $z = 0.257$ at a small projected offset ($\sim 2~$kpc). We uncover long-lived X-ray ({\it CXO}) and radio/millimeter (VLA, MeerKAT, and ALMA) afterglow emission, detected to $\sim 37~$days and $\sim 20~$days (rest frame), respectively. We measure a wide jet ($\sim 10.4^\circ$) and relatively high circumburst density ($\sim 0.07~{\rm cm}^{-3}$) compared to the short GRB population. Our data cannot be easily fit with a standard forward shock model, but they are generally well fit with the incorporation of a refreshed forward shock and a reverse shock at $< 1~$day.
     We incorporate \grb\ into a larger sample of 132 X-ray detected events, 71 of which were radio-observed (17 cm-band detections), for a systematic study of the distributions of redshifts, jet and afterglow properties, galactocentric offsets, and local environments of events with and without detected radio afterglows. 
     Compared to the entire short GRB population, the majority of radio-detected GRBs are at relatively low redshifts ($z < 0.6$) and have high circumburst densities ($> 10^{-2}~{\rm cm}^{-3}$), consistent with their smaller ($< 8~$kpc) projected galactocentric offsets. We additionally find that 70\% of short GRBs with opening angle measurements were radio-detected, indicating the importance of radio afterglows in jet measurements, especially in the cases of wide ($> 10^\circ$) jets where observational evidence of collimation may only be detectable at radio wavelengths. Owing to improved observing strategies and the emergence of sensitive radio facilities, the number of radio-detected short GRBs has quadrupled in the past decade.
\end{abstract}

\section{Introduction}

Short-duration ($< 2~$s) gamma-ray bursts (GRBs) are energetic, cosmological explosions that originate from compact object mergers involving neutron stars \citep{1989Natur.340..126E, 1992ApJ...395L..83N, 2014ARA&A..52...43B, 2017ApJ...848L..13A}. As the collimated jet produced in the explosion interacts with the surrounding medium, synchrotron emission known as the ``afterglow'' is produced \citep[e.g.,][]{1998ApJ...497L..17S, 1999ApJ...523..177W, 2002ApJ...568..820G}, which can be detected across the electromagnetic spectrum (X-rays to radio). 
The monitoring of these afterglows at multiple wavelengths not only provides precise localizations, allowing for host-galaxy associations and redshifts \citep[e.g.,][]{2022ApJ...940...56F, 2022MNRAS.515.4890O}, but also the determination of explosion properties such as energetics and environment density \citep[e.g.,][]{2001ApJ...561L.171P, 2006ApJ...650..261S, 2015ApJ...815..102F, 2020MNRAS.495.4782O}, as  well as the collimation of the relativistic jets \citep[e.g.][]{2015ApJ...815..102F, 2018ApJ...857..128J, 2020MNRAS.492.5011D, 2023ApJ...959...13R}. 

In particular, radio observations uniquely enable us to constrain the energetics and environments of short GRBs, in contrast to optical and X-ray observations which may be degenerate to these properties \citep[e.g.][]{2015ApJ...815..102F}. 
Given the importance of these observations for understanding short GRBs, dedicated campaigns have been conducted to search for the radio afterglow, resulting in the first detection in 2005 \citep{2005Natur.438..988B}.
Radio afterglow detections have also been instrumental in constraining the jet collimation of short GRBs \citep{2014ApJ...780..118F, 2016ApJ...827..102T, 2022ApJ...935L..11L, 2023arXiv230810936S}. Additionally, radio detections have uncovered afterglow behavior that deviates from the standard forward shock (FS), including energy injection and reverse shocks (RSs) \citep{2006ApJ...650..261S, 2014ApJ...780..118F, 2018Galax...6..103L, 2019ApJ...883...48L, 2019MNRAS.489.2104T,
2021ApJ...906..127F, 2023arXiv230810936S}. 

While studies of radio observations of long ($>2~$s) GRBs indicate that $\sim 30$--$60\%$ should be detectable with current radio facilities \citep{2012ApJ...746..156C,2018MNRAS.473.1512A}, in the years 2005--2015 only five ($\sim 5\%$) short GRBs were radio-detected, indicating that short GRBs typically have lower kinetic energies and reside in lower density environments than long GRBs \citep{2001ApJ...561L.171P, 2006ApJ...650..261S, 2015ApJ...815..102F, 2020MNRAS.495.4782O}.
The conclusion of the VLA upgrade in 2012 increased the sensitivity of the array by a factor of $\sim 10$ \citep{2011ApJ...739L...1P}, revolutionizing the possibility for short GRB radio afterglow discovery. Shortly thereafter, both ALMA and MeerKAT were completed (2014 and 2016, respectively), providing sensitive radio and millimeter (mm) observatories across $\sim 2$ decades of frequency. With access to these powerful observatories, in the last decade we have quadrupled the number of short GRBs detected at radio wavelengths \citep[now 16 or $\sim 10\%$;][ this work]{2005Natur.438..988B, 2006ApJ...650..261S, 2014ApJ...780..118F, 2015ApJ...815..102F, 2017GCN.21395....1F, 2019ApJ...883...48L, 2021ApJ...906..127F, 2022ApJ...935L..11L, 2023GCN.33372....1S, 2023GCN.33358....1S, 
2023arXiv230810936S, 2023GCN.35097....1R}. This includes the first detections of short GRBs for both ALMA \citep{2022ApJ...935L..11L} and MeerKAT \citep{2023arXiv230810936S}. With this larger radio-detected sample, we are now able to search for trends across redshift, afterglow properties, energetics, and circumburst and host environments, in order to understand which factors influence the radio detectability of short GRBs.

Here, we present the discovery and analysis of the multiwavelength (radio, millimeter, optical, and X-ray) afterglow of \grb, pinpointed to a host galaxy at redshift $z=0.257$.
In Section~\ref{sec:Observations_231117A}, we present the {\it Swift} burst discovery, broadband observations
including Keck afterglow spectroscopy, and host association. We also introduce unpublished VLA observations of 12~short GRBs to incorporate into a comparison sample. In Section~\ref{sec:HostGalaxy_231117A} we analyze the burst location and perform host-galaxy modeling for inferred stellar-population properties.
We model the afterglow and derive properties on the burst energetics, microphysics, and environment in Section~\ref{sec:AGModeling}. We find that an energy injection episode combined with an RS better explains the early-time ($\lesssim 1~$day) afterglow compared to a standard FS. In Section~\ref{sec:Discussion_231117A}, we take advantage of the growing radio-detected population to present the distributions of redshifts, jet and afterglow properties, and locations for bursts with and without detected radio afterglows. We conclude in Section~\ref{sec:Conclusions_231117A}. Throughout, we employ the $\Lambda$CDM cosmological parameters of H$_{0} = 68~{\rm km \, s}^{-1} \, {\rm Mpc}^{-1}$, $\Omega_{M} = 0.31$, $\Omega_{\rm \Lambda} = 0.69$ \citep{2020A&A...641A...6P}.  Also, all magnitudes are given in the AB system \citep{1983ApJ...266..713O} and are not corrected for foreground extinction.

\section{Observations}
\label{sec:Observations_231117A}

\subsection{\texorpdfstring{$\gamma$}{Gamma}-ray}

\grb was detected by the Burst Alert Telescope \citep[BAT; ][]{2005SSRv..120..143B} onboard the {\it Neil Gehrels  Swift Observatory} \citep[{\it Swift}; ][]{2004ApJ...611.1005G} at (UTC dates are used throughout this paper) 03:03:19 on 2023 November 17 \citep{2023GCN.35071....1L}. The burst duration is $T_{90}({\rm 15-150~keV}) = 0.668~$s with a fluence of $f_{\gamma} ({\rm 15-150~keV}) = 2.3\times 10^{-6}~{\rm erg~cm}^{-2}$. The position of the burst was refined to $\alpha$
= \ra{22}{09}{37.0} and $\delta =~$\dec{+13}{30}{58.1} (J2000) with a positional uncertainty of $1.0'$ radius \citep[90\% containment;][]{2023GCN.35152....1M}. Given the measured duration, \grb is classified as a short GRB. We further confirm this classification by calculating a hardness ratio (${\rm HR}$) from the BAT observations of ${\rm HR} = f_{\gamma} ({\rm 50-100~keV})/f_{\gamma} ({\rm 25-50~keV}) \approx 1.4$, placing \grb in line with other short GRBs in the hardness vs. duration plane \citep{2016ApJ...829....7L}. Furthermore, adopting $z = 0.257$ (Section~\ref{sec:lris}), we calculate an isotropic $\gamma$-ray energy of $E_{\rm iso}(15-150~{\rm keV}) \approx 3.8 \times 10^{50}~$erg and intrinsic peak energy of $E_{\rm p, i} \approx 207~$keV (assuming an observed peak energy of $E_{\rm p} \approx 165~$keV\footnote{\url{https://gcn.gsfc.nasa.gov/notices\_s/1197027/BA/}}; \citealt{2016ApJ...829....7L}), consistent with the $E_{\rm p,i }$--$E_{\rm iso}$ distribution for short GRBs \citep[e.g.,][]{2002A&A...390...81A,2008MNRAS.391..577A, 2020MNRAS.492.1919M}.

In addition to {\it Swift}, \grb also triggered a myriad of other $\gamma$-ray telescopes, including  AstroSat CZTI \citep{2023GCN.35072....1N}, AGILE \citep{2023GCN.35075....1C}, Konus-Wind \citep{2023GCN.35079....1S}, Glowbug \citep{2023GCN.35081....1C}, GRBAlpha \citep{2023GCN.35117....1X}, and CALET \citep{2023GCN.35136....1Y}. The data from all of these telescopes confirmed the short-duration nature of \grb.

\subsection{X-ray}

\subsubsection{{\it Swift} X-ray Telescope}\label{sec:swift}
 
The {\it Swift}/X-ray Telescope \citep[XRT;][]{2005SSRv..120..165B} started observations of \grb at $\delta t = 85~$s (where $\delta t$ is the time after BAT trigger) and detected an uncataloged X-ray source within the 90\% BAT region, considered to be the X-ray afterglow of \grb \citep{2023GCN.35071....1L, 2023GCN.35128....1M}. The XRT-detected afterglow has a refined position of $\alpha =~$\ra{22}{09}{33.57} and $\delta =~$\dec{+13}{31}{21.1} (J2000), and a positional uncertainty of $2.0''$ radius \citep[90\% containment;][]{2009MNRAS.397.1177E, 2023GCN.35074....1B}. {\it Swift} continued to observe and detected the X-ray afterglow of \grb until $\delta t \approx 11.9~$days. To create a more sampled XRT afterglow than the light curve available in the repository, we used the rebinning tool in the \swift\ light-curve repository\footnote{\url{https://www.swift.ac.uk/xrt_curves/01197027/}}. We chose the dynamic binning option, and set the rate factor and minimum counts per bin to 5 in order to obtain more data points in the light curve, particularly during the plateau phase (see below).
The XRT afterglow light curve of \grb shows a complex structure in which an initial afterglow steep-decay phase is followed by a short flare at $\delta t \approx 0.06~$days. Right after this event, the afterglow exhibits a plateau phase that extends from $\delta t \approx 0.2~$--$0.8~$days, and is continued by the final afterglow decay phase at $\delta t \gtrsim 2.4~$days. 

\subsubsection{{\it Chandra} X-ray Observatory}\label{sec:chandra}

We initiated observations of \grb\ with the Advanced CCD Imaging Spectrometer detector (ACIS; \citealt{2003SPIE.4851...28G}) onboard the {\it Chandra X-ray Observatory} \citep[{\it CXO}; ][]{2000SPIE.4012....2W} starting at $\delta t = 4.3~$ days, under Program 24400307 (PI W. Fong; ObsIDs 265[46-48], 291[13-14], and 291[58-61]), and divided across three epochs. We reduced and analyzed the {\it CXO} observations using the \texttt{CIAO} software package \citep[v.\,4.12;][]{2006SPIE.6270E..1VF} with calibration files (\texttt{caldb}; v.\,4.9.0). New Level II event files were obtained after reprocessing the data utilizing \texttt{chandra\_repro}, and filtered the dataset against any high background activity. The absolute astrometry was corrected by running the corresponding \texttt{CIAO} task\footnote{\url{https://cxc.cfa.harvard.edu/ciao/threads/reproject\_aspect/}} using the USBO-B1.0 optical catalog. We then performed a blind search for X-ray sources with \texttt{CIAO}/\texttt{wavdetect} on the first {\it CXO} epoch (ObsID 26546, $\sim 27$\,ks effective exposure time) at $\delta t \approx 4.3~$days. The afterglow of \grb is detected at a position of $\alpha =~$\ra{22}{09}{33.37}, $\delta =~$\dec{+13}{31}{20.0} (J2000) with a total positional uncertainty of $0.6\arcsec$ ($1 \sigma$). We note that the {\it CXO} afterglow region does not coincide with the 90\% confidence XRT region; indeed, the centers of these regions are located 3.1\arcsec\ from each other. Owing to {\it CXO} observing constraints, the second and third epochs were taken with multiple exposures ($\delta t \approx 25.6$~days, ObsIDs 26547, 29113, and 29114, effective exposure time of $\sim 36.5$\,ks; $\delta t \approx 46.5$~days, ObsIDs 26548 and 291[58-61], effective exposure time of $\sim 53$\,ks). We therefore combined the exposures using \texttt{CIAO}/\texttt{merge\_obs} to generate a single observation per epoch and continued to detect the X-ray afterglow through the final epoch. At $z = 0.257$ (Section~\ref{sec:lris}), \grb is detected in the X-ray band out to a rest-frame time of $\delta t_{\rm rest} \approx 37~$days, and thus constitutes the longest detected X-ray afterglow of a short GRB to date \citep[the previous record holder was GRB\,150101B, at $\delta t_{\rm rest} \approx 35~$days;][]{2016ApJ...833..151F}. A summary of our {\it CXO} observations is presented in Table~\ref{tab:data_xray_231117A}.

\subsubsection{Spectral Analysis}\label{sec:spectral_analysis_xray}
We use the ``create time-sliced spectra'' option\footnote{\url{https://www.swift.ac.uk/xrt_spectra/01197027/}} in the \swift\ repository to generate the spectra ($0.3-10~$keV) for each new time bin of the XRT afterglow light curve of \grb. In order to obtain the {\it CXO} spectra, we use a circular region with a radius of 3.0\arcsec\ centered on the {\it CXO} afterglow position, and obtain the background from a source-free annulus with inner and outer radii of 15\arcsec\ and 30\arcsec, respectively. We generate the source and background spectra for the individual {\it CXO} observations of the \grb afterglow, as well as the necessary ancillary response file (\texttt{arf}) and redistribution matrix file (\texttt{rmf}) utilizing the \texttt{CIAO}/\texttt{specextract} tool.

We use \texttt{Xspec} (v.12.9.0; \citealt{1996ASPC..101...17A}) to perform the spectral fitting within the $0.5-8$\,keV energy band. We choose a bin size with at least one count per bin, so that we avoid any bin with negative net values after the background subtraction. In addition, we set the abundances to \texttt{WILM} (\citealt{2000ApJ...542..914W}), the X-ray cross-sections to \texttt{VERN} (\citealt{1996ApJ...465..487V}), and the statistics to W-statistics (statistics for background-subtracted Poisson data; \citealt{1979ApJ...230..274W}). No spectral evolution is seen between observations at $\delta t \gtrsim 3~$days, so we jointly fit the \swift\ and {\it CXO} spectra using a single-absorbed power-law model (\texttt{[const x tbabs x ztbabs x pow]$_{\text{AG}}$}). We set the cross-calibration constants between {\it CXO}/ACIS-S3 (High Resolution Imaging mode) and \swift/XRT-PC to 1 and 0.872, respectively \citep[Table 5;][]{2017A&A...597A..35P}. For fitting our model, we fix the Galactic contribution to $N_{{\rm{H, MW}}}=6.1\times10^{20}$\,cm$^{-2}$ \citep{2013MNRAS.431..394W} and the redshift to $z = 0.257$ (see Section~\ref{sec:lris}). No evidence for intrinsic absorption  ($N_{\rm H, int}$) is found, so we fix $N_{\rm H, int} = 0$ for our analysis. We determine the X-ray photon index ($\Gamma_{\rm X}$) of the joint spectral fitting to be $\Gamma_{\rm X} = 1.46\pm 0.12$ (1$\sigma$) for $\delta t \gtrsim 1~$days, corresponding to an X-ray spectral index of $\beta_{\rm X} = -0.46 \pm 0.12$ ($\beta_{\rm X} \equiv 1-\Gamma_{\rm X}$, $F_{\nu} \propto \nu^{\beta}$; uncertainties are 1$\sigma$). Finally, we calculate unabsorbed fluxes ($0.3-10$\,keV) using \texttt{Xspec cflux}. The unabsorbed fluxes with $1\sigma$ uncertainties and flux density at 1\,keV for subsequent analysis are listed in Table\,\ref{tab:data_xray_231117A}.

\begin{deluxetable*}{cccc}
 \tabletypesize{\footnotesize}
 \tablecolumns{4}
 \tablecaption{X-ray observations of \grb\label{tab:xray_spectra}}
 \tablewidth{5pt}
 \tablehead{   
   \colhead{$\delta t^{\rm{a}}$} &
   \colhead{Bin length}&
   \colhead{Unabsorbed flux (0.3--10~keV)$^{\rm{b}}$}&
   \colhead{Flux Density at 1~keV$^{\rm{b}}$}\\
   \colhead{(days)} &
   \colhead{(s)} &
   \colhead{($10^{-14}$\,erg\,s$^{-1}$\,cm$^{-2}$)} &
   \colhead{($\mu$Jy)}
   }
\startdata 
\multicolumn{4}{c}{{\it Swift}/XRT} \\
\hline
$1.4 \times 10 ^{-3}$  &  $1.5\times 10 ^{1}$   &   $(9.7 \pm 3.0) \times 10^{3}$  & $(1.0 \pm 0.3)\times 10^{-5}$ \\
$1.6 \times 10 ^{-3}$  &  $1.8\times 10 ^{1}$   &   $(7.4 \pm 2.3) \times 10^{3}$  & $(7.7 \pm 2.5)\times 10^{-6}$ \\
$1.7 \times 10 ^{-3}$  &  $1.0\times 10 ^{1}$   &   $(1.5 \pm 0.4) \times 10^{4}$  & $(1.5 \pm 0.5)\times 10^{-5}$ \\
$1.9 \times 10 ^{-3}$  &  $1.8\times 10 ^{1}$   &   $(8.2 \pm 2.5) \times 10^{3}$  & $(8.6 \pm 2.6)\times 10^{-6}$ \\
$2.1 \times 10 ^{-3}$  &  $1.3\times 10 ^{1}$   &   $(1.2 \pm 0.4) \times 10^{4}$  & $(1.3 \pm 0.4)\times 10^{-5}$ \\
$2.2 \times 10 ^{-3}$  &  $1.8\times 10 ^{1}$   &   $(7.8 \pm 2.4) \times 10^{3}$  & $(8.2 \pm 2.5)\times 10^{-6}$ \\
$2.6 \times 10 ^{-3}$  &  $3.3\times 10 ^{1}$   &   $(5.3 \pm 1.5) \times 10^{3}$  & $(5.5 \pm 1.5)\times 10^{-6}$ \\
$2.9 \times 10 ^{-3}$  &  $4.3\times 10 ^{1}$   &   $(4.0 \pm 1.1) \times 10^{3}$  & $(4.2 \pm 1.2)\times 10^{-6}$ \\
$3.4 \times 10 ^{-3}$  &  $3.0\times 10 ^{1}$   &   $(5.2 \pm 1.5) \times 10^{3}$  & $(5.4 \pm 1.6)\times 10^{-6}$ \\
$4.0 \times 10 ^{-3}$  &  $8.5\times 10 ^{1}$   &   $(1.5 \pm 0.5) \times 10^{3}$  & $(1.6 \pm 0.5)\times 10^{-6}$ \\
$5.0 \times 10 ^{-3}$  &  $8.8\times 10 ^{1}$   &   $(6.4 \pm 2.0) \times 10^{2}$  & $(6.6 \pm 2.1)\times 10^{-7}$ \\
$5.9 \times 10 ^{-3}$  &  $5.8\times 10 ^{1}$   &   $(1.0 \pm 0.3) \times 10^{3}$  & $(1.0 \pm 0.3)\times 10^{-6}$ \\
$6.7 \times 10 ^{-3}$  &  $9.0\times 10 ^{1}$   &   $(6.3 \pm 2.0) \times 10^{2}$  & $(6.6 \pm 2.1)\times 10^{-7}$ \\
$7.6 \times 10 ^{-3}$  &  $4.5\times 10 ^{1}$   &   $(1.3 \pm 0.4) \times 10^{3}$  & $(1.3 \pm 0.4)\times 10^{-6}$ \\
$8.1 \times 10 ^{-3}$  &  $5.3\times 10 ^{1}$   &   $(1.1 \pm 0.3) \times 10^{3}$  & $(1.1 \pm 0.4)\times 10^{-6}$ \\
$8.8 \times 10 ^{-3}$  &  $5.5\times 10 ^{1}$   &   $(1.0 \pm 0.3) \times 10^{3}$  & $(1.1 \pm 0.3)\times 10^{-6}$ \\
$9.3 \times 10 ^{-3}$  &  $5.8\times 10 ^{1}$   &   $(1.1 \pm 0.3) \times 10^{3}$  & $(1.1 \pm 0.3)\times 10^{-6}$ \\
$1.0 \times 10 ^{-2}$  &  $6.0\times 10 ^{1}$   &   $(9.6 \pm 3.0) \times 10^{2}$  & $(1.0 \pm 0.3)\times 10^{-6}$ \\
$1.1 \times 10 ^{-2}$  &  $7.8\times 10 ^{1}$   &   $(7.0 \pm 2.3) \times 10^{2}$  & $(7.3 \pm 2.4)\times 10^{-7}$ \\
$1.2 \times 10 ^{-2}$  &  $8.3\times 10 ^{1}$   &   $(6.6 \pm 2.1) \times 10^{2}$  & $(6.9 \pm 2.2)\times 10^{-7}$ \\
$1.3 \times 10 ^{-2}$  &  $1.0\times 10 ^{2}$   &   $(5.2 \pm 1.7) \times 10^{2}$  & $(5.5 \pm 1.8)\times 10^{-7}$ \\
$1.4 \times 10 ^{-2}$  &  $8.0\times 10 ^{1}$   &   $(7.1 \pm 2.3) \times 10^{2}$  & $(7.4 \pm 2.4)\times 10^{-7}$ \\
$1.5 \times 10 ^{-2}$  &  $1.3\times 10 ^{2}$   &   $(4.1 \pm 1.3) \times 10^{2}$  & $(4.3 \pm 1.4)\times 10^{-7}$ \\
$1.6 \times 10 ^{-2}$  &  $6.0\times 10 ^{1}$   &   $(9.1 \pm 2.9) \times 10^{2}$  & $(9.5 \pm 3.0)\times 10^{-7}$ \\
$1.7 \times 10 ^{-2}$  &  $7.5\times 10 ^{1}$   &   $(7.2 \pm 2.3) \times 10^{2}$  & $(7.5 \pm 2.4)\times 10^{-7}$ \\
$1.8 \times 10 ^{-2}$  &  $4.5\times 10 ^{1}$   &   $(1.2 \pm 0.4) \times 10^{3}$  & $(1.3 \pm 0.4)\times 10^{-6}$ \\
$1.9 \times 10 ^{-2}$  &  $1.6\times 10 ^{2}$   &   $(4.5 \pm 1.3) \times 10^{2}$  & $(4.7 \pm 1.3)\times 10^{-7}$ \\
$5.8 \times 10 ^{-2}$  &  $8.5\times 10 ^{1}$   &   $(6.3 \pm 2.0) \times 10^{2}$  & $(6.6 \pm 2.1)\times 10^{-7}$ \\
$5.9 \times 10 ^{-2}$  &  $6.3\times 10 ^{1}$   &   $(8.7 \pm 2.8) \times 10^{2}$  & $(9.1 \pm 2.9)\times 10^{-7}$ \\
$6.0 \times 10 ^{-2}$  &  $2.5\times 10 ^{1}$   &   $(2.2 \pm 0.7) \times 10^{3}$  & $(2.3 \pm 0.7)\times 10^{-6}$ \\
$6.1 \times 10 ^{-2}$  &  $6.3\times 10 ^{1}$   &   $(8.7 \pm 2.8) \times 10^{2}$  & $(9.0 \pm 2.9)\times 10^{-7}$ \\
$6.2 \times 10 ^{-2}$  &  $1.5\times 10 ^{2}$   &   $(3.7 \pm 1.2) \times 10^{2}$  & $(3.9 \pm 1.2)\times 10^{-7}$ \\
$6.4 \times 10 ^{-2}$  &  $1.8\times 10 ^{2}$   &   $(2.9 \pm 1.0) \times 10^{2}$  & $(3.0 \pm 1.0)\times 10^{-7}$ \\
$6.5 \times 10 ^{-2}$  &  $3.8\times 10 ^{1}$   &   $(1.5 \pm 0.5) \times 10^{3}$  & $(1.5 \pm 0.5)\times 10^{-6}$ \\
$6.6 \times 10 ^{-2}$  &  $9.3\times 10 ^{1}$   &   $(5.8 \pm 1.9) \times 10^{2}$  & $(6.1 \pm 2.0)\times 10^{-7}$ \\
$6.7 \times 10 ^{-2}$  &  $1.6\times 10 ^{2}$   &   $(4.8 \pm 1.3) \times 10^{2}$  & $(5.0 \pm 1.4)\times 10^{-7}$ \\
$2.1 \times 10 ^{-1}$  &  $9.4\times 10 ^{2}$   &   $(6.8 \pm 2.3) \times 10^{1}$  & $(7.1 \pm 2.3)\times 10^{-8}$ \\
$7.9 \times 10 ^{-1}$  &  $8.0\times 10 ^{2}$   &   $(7.2 \pm 2.4) \times 10^{1}$  & $(7.6 \pm 2.5)\times 10^{-8}$ \\
$2.4$                  &  $1.7\times 10 ^{3}$   &   $(3.5 \pm 1.1) \times 10^{1}$  & $(3.7 \pm 1.2)\times 10^{-8}$ \\
$2.5$                  &  $6.6\times 10 ^{3}$   &   $(2.9 \pm 1.0) \times 10^{1}$  & $(3.1 \pm 1.0)\times 10^{-8}$ \\
$2.6$                  &  $1.2\times 10 ^{3}$   &   $(5.8 \pm 1.8) \times 10^{1}$  & $(6.1 \pm 1.8)\times 10^{-8}$ \\
$3.7$                  &  $2.5\times 10 ^{4}$   &   $(1.9 \pm 0.6) \times 10^{1}$  & $(2.0 \pm 0.6)\times 10^{-8}$ \\
$4.1$                  &  $7.6\times 10 ^{4}$   &   $(3.1 \pm 0.8) \times 10^{1}$  & $(3.2 \pm 0.8)\times 10^{-8}$ \\
$5.7$                  &  $1.3\times 10 ^{4}$   &   $(3.3 \pm 0.8) \times 10^{1}$  & $(3.5 \pm 0.9)\times 10^{-8}$ \\
$7.4$                  &  $2.2\times 10 ^{5}$   &     $7.3 \pm 2.4$                & $(7.6 \pm 2.5)\times 10^{-9}$ \\
$10.3$                 &  $1.8\times 10 ^{5}$   &     $8.5 \pm 2.5$                & $(8.9 \pm 2.6)\times 10^{-9}$ \\
\hline
\multicolumn{4}{c}{{\it CXO}/ACIS-S} \\
\hline
$4.3$             &  $2.7\times 10 ^{4}$   &  $19.7 \pm 1.2$        & $(2.1 \pm 0.1)\times 10^{-8}$ \\
$25.5^{\rm{c}}$   &  $3.6\times 10 ^{4}$   &  $1.2_{-0.3}^{+0.4}$   & $(1.2 \pm 0.4)\times 10^{-9}$ \\
$46.5^{\rm{d}}$   &  $5.3\times 10 ^{4}$   &  $0.3 \pm 0.2$         & $(2.6 \pm 1.7)\times 10^{-10}$ \\
\enddata
\tablecomments{
$^{\rm{a}}$ Time is log-centered (observer frame). \\ $^{\rm{b}}$ Uncertainties correspond to $1\sigma$ confidence. \\ $^{\rm{c}}$ Second {\it CXO} epoch combining ObsIDs 26547, 291[13-14]. \\ $^{\rm{d}}$ Third {\it CXO} epoch combining ObsIDs 26548, 291[58-61]. 
}
\label{tab:data_xray_231117A}
\end{deluxetable*}

\subsection{Optical and Near-Infrared}\label{sec:optical}

The optical afterglow of \grb was observed by numerous telescopes on  timescales of hours (first reported by \citealt{2023GCN.35083....1Y} at $\delta t = 0.332~$days) to weeks following the initial GRB trigger. Here we describe our optical through near-infrared (IR) spectroscopic and photometric follow-up observations, including data on  the host galaxy. We list our optical afterglow and host-galaxy photometry in Table~\ref{tab:optical}.

\subsubsection{Keck/LRIS}\label{sec:lris}

\begin{figure*}
    \includegraphics[width=\textwidth]{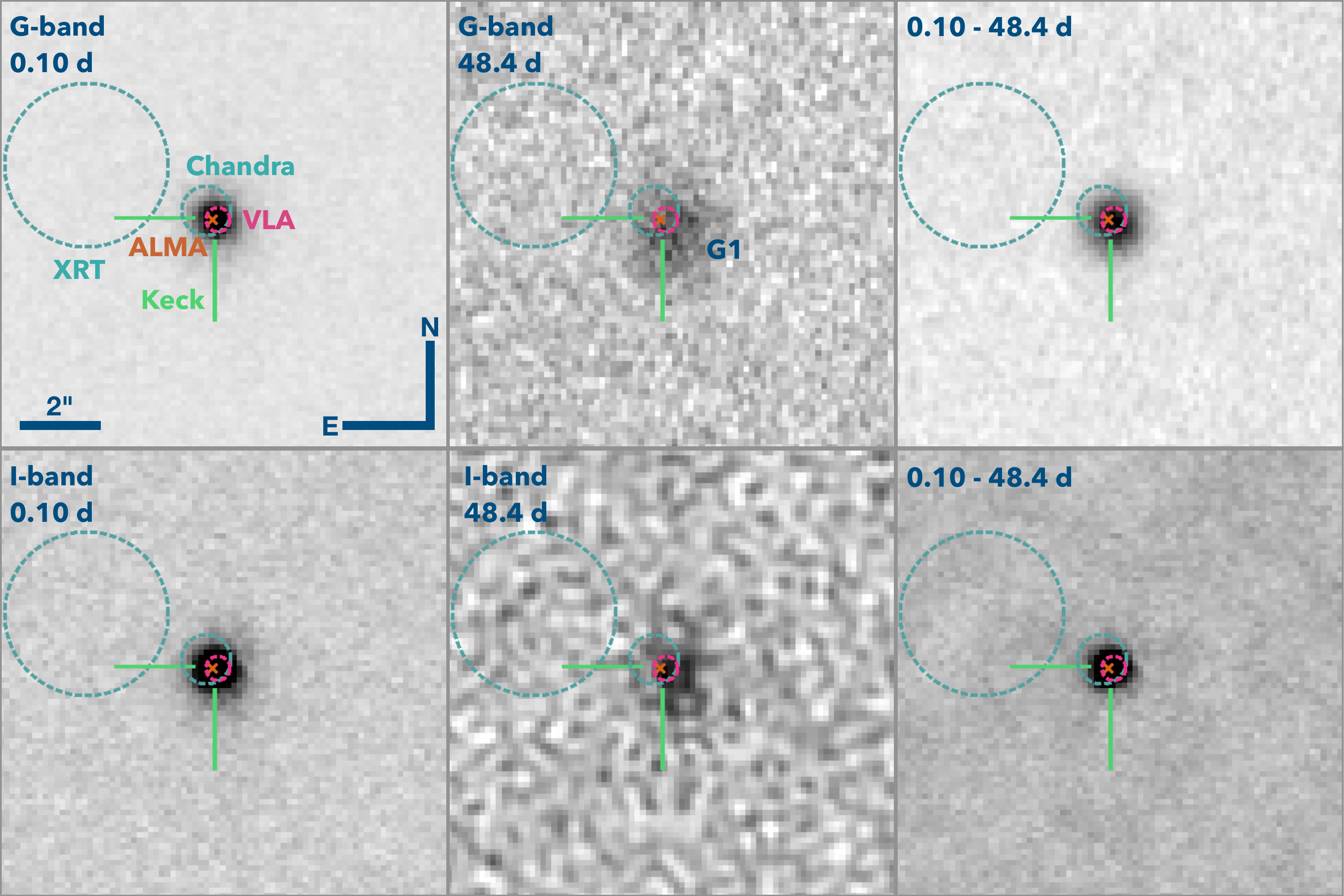}
    \caption{Keck/LRIS $G$- and $I$-band imaging of \grb\ (left panels) obtained on 2023 November 17 and 2024 January 4, as described in Section~\ref{sec:lris}.  The template image highlights the location of the \grb\ host galaxy ``G1'' (middle panels).  We also show our subtraction between the two epochs in both bands (right panels), demonstrating that \grb\ is clearly detected at high significance in the first epoch.  For comparison to these data, we highlight the positions of the {\it Swift}/XRT and {\it CXO} X-ray counterpart localizations (blue circles), the Keck/LRIS optical localization derived from these difference images (green lines), the ALMA millimeter counterpart localization (orange cross), and the VLA radio counterpart localization (red circle).}\label{fig:difference}
\end{figure*}

At 05:22:15.62 on 2023 November 17 ($\delta t \approx 0.1~$days), we observed the location of GRB\,231117A with the Low Resolution Imaging Spectrometer (LRIS; \citealt{1995PASP..107..375O}) mounted on Keck I (Program U192; PI R. Margutti). We obtained $6 \times 150$~s of imaging in each of the $G$- and $I$-band filters. The images were processed and coadded using a custom {\tt POTPyRI} Python package\footnote{https://github.com/CIERA-Transients/POTPyRI}, which applies bias and flat-fielding corrections using calibration frames obtained on the same night and instrumental configuration, image alignment and registration to the {\it Gaia} DR3 astrometric frame \citep{2023A&A...674A...1G}, and image stacking using {\tt SWarp} \citep{2010ascl.soft10068B}. We noted the presence of an optical source with $G=20.3\pm0.1$~mag and $I=19.8\pm0.1$~mag \citep[see also][]{2023GCN.35087....1R}, the counterpart of \grb, that is deeply embedded in a moderately faint ($r_{\rm PS1}=21.58$~mag) galaxy (``G1'' in  Figure~\ref{fig:difference}), resulting in significant blending between the optical counterpart and G1 in the first epoch of imaging. Therefore, we reobserved the same field with Keck/LRIS $G$+$I$ bands on 2024 January 4 ($\delta t \approx 48.4~$days) and detected no signature of \grb\ when compared with G1's catalogud brightness (Section~\ref{sec:PS1}). We therefore performed image subtraction between the two epochs using {\tt hotpants} \citep{2015ascl.soft04004B} with default parameters for each frame.  The optical counterpart to \grb\ is detected in both bands and aperture photometry is performed with {\tt DoPhot} \citep{1993PASP..105.1342S}. We show our Keck/LRIS difference imaging in Figure~\ref{fig:difference}, highlighting the detection of \grb. 

In order to measure the absolute position of \grb, we first quantify the precision of our alignment between the {\it Gaia} DR3 astrometric frame and our first epoch of Keck/LRIS $G$-band imaging.  There are 89 {\it Gaia} astrometric standards within the roughly 5.7$\times$8.1~arcmin$^{2}$ field of view of the Keck/LRIS image. We calculate an astrometric solution for the LRIS image frame from these sources using the {\tt IRAF}\footnote{IRAF is distributed by the National Optical Astronomy Observatory, which is operated by the Association of Universities for Research in Astronomy (AURA) under a cooperative agreement
with the National Science Foundation.} task {\tt ccmap} to a tangent plane projection along with a fourth-order polynomial distortion solution \citep{1986SPIE..627..733T}.  This results in an absolute alignment uncertainty of 0.074\arcsec\ in right ascension and 0.052\arcsec\ in declination.  After performing the subtraction procedure with our template image, we use {\tt sextractor} to derive a position for the afterglow of $\alpha =~$\ra{22}{09}{33.3554} $\pm 0.0005$ and $\delta =~$\dec{+13}{31}{19.809} $\pm 0.004$ (J2000). The resulting optical afterglow location is consistent with our measured {\it CXO} localization (Section~\ref{sec:chandra}), and is similarly offset from the XRT localization.  Given the spatial coincidence between the afterglow and galaxy, and lack of any other clear host candidates, we consider G1 to be the host of \grb.

\begin{figure}
   \centering
    \includegraphics[width = 0.48\textwidth]{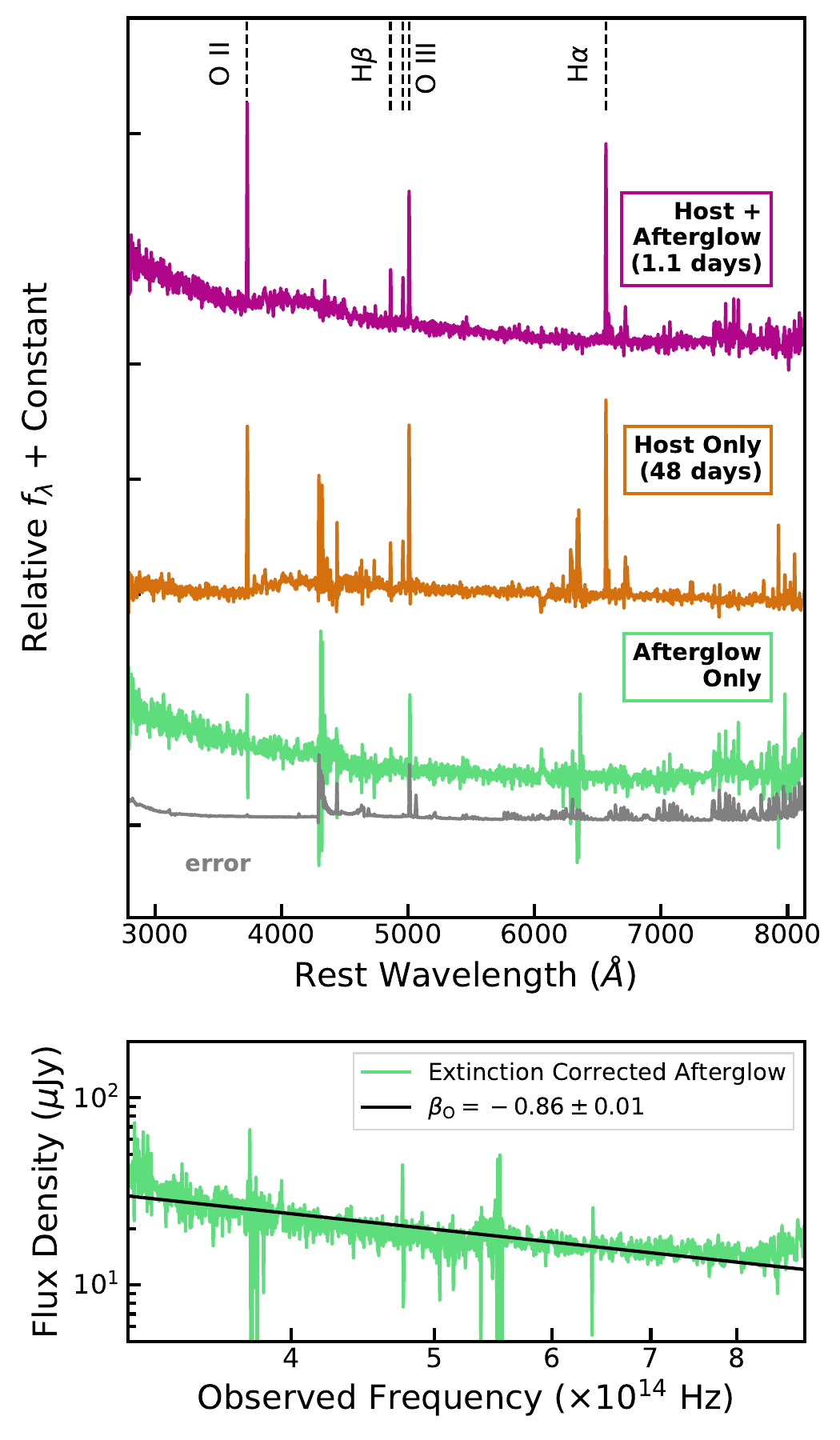}
   \caption{{\emph Top:} LRIS spectra of the location of \grb taken at $\delta t \approx 1.1~$days (``Host $+$ Afterglow'') and $\delta t \approx 48~$days (``Host Only''). We identify the locations of the emission lines that we used to determine the redshift of \grb at $z = 0.257$. Also displayed is the residual spectrum from subtracting our LRIS Host Only spectrum from our Host $+$ Afterglow spectrum (see subtraction procedure in Section~\ref{sec:ag-subtraction}).  The resulting ``Afterglow Only'' spectrum is shown as a green curve with the uncertainty spectrum, including uncertainty in both spectral extractions and the relative scaling between the spectra, shown in gray.\\
   {\emph Bottom:} The Afterglow Only spectrum converted to flux density vs. frequency space, and corrected for Milky Way extinction \citep{1989ApJ...345..245C}. We fit a power law to the afterglow spectrum (black line) and determine $\beta_{\rm O} = -0.86 \pm 0.01$ ($F_\nu \propto \nu^{\beta}$).}\label{fig:afterglow}
\end{figure}

At 05:03:50.4 on  2023 November 18 ($\delta t \approx 1.1~$days), we obtained a spectrum with Keck/LRIS (exposure times of $2 \times 600$~s on both the blue and red sides; Program U127; PI M. Malkan). The spectrum was obtained with the 1.0\arcsec\ slit and the D560 dichroic with the 600/4000 grism on the blue side and the 400/8500 grating on the red side, covering the wavelength range  $\sim 3140$--10200~\AA. We perform overscan subtraction, flat-fielding, background subtraction, wavelength calibration, and flux calibration using a custom series of IDL routines.
At this epoch, both light from the afterglow and host galaxy contribute. We coadd the one-dimensional (1D) spectra and confirm the presence of the [O~\textsc{ii}] $\lambda\lambda$3726, 3729 doublet, H$\beta$ $\lambda4861$, [O~\textsc{iii}]$\lambda5007$, and H$\alpha$ $\lambda6563$ in emission at $z=0.257$ (Figure~\ref{fig:afterglow}). The presence of multiple emission lines, coupled with lack of strong 4000~\AA break or any absorption lines, indicates the host of \grb\ is a star-forming galaxy.

Finally, at 13:35:33 on 2024 January 4 ($\delta t \approx 48.4~$days), we obtained a spectrum of G1 with Keck/LRIS ($3\times1200$~s on the blue side, and $2\times1200$~s on the red side). The spectrum was obtained with the 1.0\arcsec\ slit and the 560 dichroic with the 400/3400 grism on the blue side and the 400/8500 grating on the red side, covering the wavelength range  $\sim 3000$--10100~\AA. We perform overscan subtraction, flat-fielding, background subtraction, and wavelength calibration using the HgNeArCdZnKrXe lamps with the Python Spectroscopic Data Reduction Pipeline (PypeIt; \citealt{2020zndo...3743493P}). The 1D object and uncertainty spectrum are extracted using the \texttt{boxcar} method with a 1.5\arcsec\ radius to encompass the full trace. We determine the flux calibration using a spectrum of the standard star Feige 34 taken on the same night.

\subsubsection{SOAR/Goodman}\label{sec:soar}

We observed the location of \grb\ using the Goodman High-throughput Spectrograph \citep{2004SPIE.5492..331C} on the SOAR 4.1\,m telescope at Cerro Pach\'on, Chile, over 5 epochs in the range 18--23 Nov. 2023 ($\delta t \approx 0.9$--$5.9~$days).  We used the Goodman imaging mode in $2\times 2$ binning (resulting in a scale of $0.3''$ per pixel), with the SDSS-$gri$ filters as detailed in Table~\ref{tab:optical}.  We processed all imaging using bias and dome-flat frames using a custom-built {\tt photpipe} pipeline \citep[see][]{2005ApJ...634.1103R,2023ApJ...954...80G}, including corrections for bias and flat-fielding, astrometric correction using {\it Gaia} DR3 standards \citep{2023A&A...674A...1G}, masking, optimal image stacking with {\tt SWarp} \citep{2010ascl.soft10068B}, point-spread function (PSF) photometry with {\tt DoPhot} \citep{1993PASP..105.1342S}, and calibration using Pan-STARRS DR2 $gri$ photometric standard stars \citep{2020ApJS..251....7F}.

In order to perform difference imaging following the same procedures as for the Keck/LRIS imaging in Section~\ref{sec:lris}, we obtained the deepest possible template image frames from before the GRB using DES Legacy images in the $gr$ bands \citep{2019AJ....157..168D} and Pan-STARRS 3$\pi$ images in the $i$ band \citep{2016arXiv161205560C,2020ApJS..251....3M}, as we could not obtain sufficiently deep SOAR templates before the source set.  We subtract these images from our SOAR/Goodman imaging using {\tt hotpants} \citep{2015ascl.soft04004B} and performed forced photometry at the site of the optical counterpart using a custom version of {\tt dophot}.  The resulting photometry of the optical afterglow is presented in Table~\ref{tab:optical}.

\begin{deluxetable}
{lcccc}
\tabletypesize{\scriptsize}
\tablecaption{Optical Photometry of the Afterglow Counterpart and Host Galaxy of \grb}
\tablewidth{5pt}
\tablehead{
\colhead{Observatory} & 
\colhead{$\delta t^{a}$} & 
\colhead{Exposure} & 
\colhead{Band} & 
\colhead{Magnitude$^{b}$} \\
&
(days) &
(s) &
&
(AB mag)
}
\startdata
\multicolumn{5}{c}{Afterglow Photometry} \\
\hline
Keck & 0.1 & 6$\times$150 & $G$ & $20.18 \pm 0.02$ \\
 & 0.1 & 6$\times$150 & $I$ & $20.25 \pm 0.05$ \\
\hline
SOAR & 0.91 & 1$\times$200 & $r$ & $20.93 \pm 0.03$ \\
 & 1.92 & 1$\times$900 & $r$ & $21.76 \pm 0.03$ \\
 & 3.93 & 3$\times$200 & $r$ & $23.07 \pm 0.19$ \\
 & 3.93 & 3$\times$200 & $i$ & $23.55 \pm 0.31$ \\
 & 4.92 & 6$\times$200 & $g$ & $\lesssim 23.8$ \\
 & 5.91 & 3$\times$200 & $r$ & $\lesssim 23.6$ \\
 & 5.92 & 3$\times$200 & $i$ & $\lesssim 22.41$ \\
\hline
T80S & 1.92 & 2$\times$600 & $r$ & $21.85 \pm 0.21$ \\
 & 1.93 & 2$\times$600 & $i$ & $\lesssim 21.54$ \\
 & 3.91 & 4$\times$600 & $r$ & $\lesssim 21.72$ \\
 & 3.94 & 3$\times$600 & $i$ & $\lesssim 21.42$ \\
\hline
\multicolumn{5}{c}{Host-Galaxy Photometry} \\
\hline
Pan-STARRS & -- & $17\times43$ & $g$ & $22.17 \pm 0.08$ \\
 & -- & $17\times40$ & $r$ & $21.74 \pm 0.07$ \\
 & -- & $37\times45$ & $i$ & $21.35 \pm 0.06$ \\
 & -- & $21\times30$ & $z$ & $21.22 \pm 0.13$ \\
\hline
MMIRS & $17.01$ & $42\times60$ & $J$ & $20.98 \pm 0.06$ \\
  &  $16.02$ & $10\times65$ & $K$ & $20.89 \pm 0.17$ \\
\enddata
\tablecomments{\emph{Top:} Our optical photometry of the counterpart to \grb\ derived using difference imaging and described in Section~\ref{sec:optical}.
\emph{Bottom:} Optical and near-infrared photometry for the host of \grb, used in the \texttt{Prospector} SED fitting in Section \ref{sec:prospector}. \\
$^{a}$ Time since \swift/BAT trigger.\\
$^{b}$ Magnitudes are not corrected for Galactic extinction.}
\end{deluxetable}
\label{tab:optical}

\subsubsection{T80-South Telescope}

We observed \grb\ with the T80-South telescope at the Cerro Tololo Inter-American Observatory, Chile over 2 epochs between 19 and 21 November 2023 ($\delta t \approx 1.9$--$3.9~$days) in the $r$ and $i$ bands.  The observations were processed  following similar procedures to those described in Section~\ref{sec:soar} for the SOAR/Goodman telescope and as described in detail by \citet{2024MNRAS.529...59S}.  We obtained template observations with the same telescope system on 2023 December 1 in the $i$ band for $3 \times 300$~s and on 2023 December 2 in $r$-band for $4\times 300$~s.  Finally, we subtract these template images from the science observations using {\tt hotpants}.  We detect a residual in our first epoch of $r$-band imaging at $\delta t \approx 1.92$~days post-burst with $r_{\rm T80S}=21.85\pm0.21$~mag, but we did not obtain a detection in any other T80S image. We instead derive 3$\sigma$ upper limits for other epochs, and report these in Table~\ref{tab:optical}.

\subsubsection{Archival Pan-STARRS Imaging}
\label{sec:PS1}

We further investigated the host galaxy of \grb\ using archival $grizY$ Pan-STARRS imaging \citep{2020ApJS..251....3M,2020ApJS..251....4W}.  We downloaded stacked and calibrated cutouts of the field centered on the site of \grb\ and estimate the brightness of the host galaxy using {\tt photutils}-based aperture photometry with a 4\arcsec\ radius aperture centered at $\alpha =~$\ra{22}{09}{33.326} and $\delta =~$\dec{+13}{31}{19.47} (J2000) and a background annulus with inner and outer radii 8\arcsec\ and 16\arcsec, respectively.  The host galaxy is clearly detected in  $griz$ bands, but  there is no  detection in $Y$ so we do not use this band in the host-galaxy analysis described below (Section~\ref{sec:prospector}).

\subsubsection{MMT/MMIRS}

We imaged the location of the host galaxy of \grb\ with the MMT and Magellan Infrared Spectrograph (MMIRS; \citealt{2012PASP..124.1318M}) mounted on the MMT on 2023 December 3--4 ($\delta t \approx 16.02$--$17.01~$days). We obtained $42\times60$~s in the $J$ band (Program 2023B-UAO-G205-23B; PI Rastinejad) and $10\times65$~s in the $K$ band (Program 2023B-UAO-S151-23B; PI Shrestha).
These images were reduced with the same procedure as was used for Keck/LRIS imaging (Section~\ref{sec:lris}). We do not expect contamination from the GRB counterpart in our host photometry, given the event redshift as well as the expected brightness of any afterglow or kilonova emission at this epoch \citep[as in, e.g.,][]{2021ApJ...916...89R}. Thus, we perform aperture photometry with IRAF/\texttt{phot} on the host galaxy (Table~\ref{tab:optical}), calibrating to 2MASS stars in the field \citep{2006AJ....131.1163S}.

\subsubsection{Additional Optical Observations}
\label{sec:Additional_Optical}

For comparison, we gather afterglow observations from the literature
\citep{2023GCN.35083....1Y, 2023GCN.35089....1K, 2023GCN.35105....1C, 2023GCN.35121....1F, 2023GCN.35124....1S, 2023GCN.35172....1P} in $g$, $r$, $r'$, $R$, and $i$. Given the location of \grb in its host, we only consider observations for which preliminary host subtraction has been performed. We use these observations to compare to our afterglow model in Section~\ref{sec:AGModeling}, but we do not include them in the modeling owing to inhomogeneities in how host subtraction was performed.

\subsection{Radio and Millimeter Afterglow Discovery and Follow-up Observations}

\subsubsection{Very Large Array}
\label{sec:VLA_231117A}

A radio source coincident with the XRT and optical afterglow positions was detected with the Australia Telescope Compact Array (ATCA) with a 9.0 GHz flux density of $F_{\nu} \approx 210~\mu$Jy at a midtime of $\delta t \approx 0.54~$days \citep{2023GCN.35097....1R}. Soon after, we initiated X-band (7.98--12.02~GHz, central frequency of 10.0~GHz) observations of \grb with the Karl G. Jansky Very Large Array (VLA) at a midtime of $\delta t = 0.87~$days (Program 23A-296; PI Schroeder). We used 3C48 for flux and bandpass calibration and J2152+1734 for complex gain calibration. The data were reduced using the Common Astronomy Software Applications \citep[\texttt{CASA};][]{2007ASPC..376..127M, 2022PASP..134k4502V} VLA Pipeline, and self-calibrated the data using the \texttt{CASA} Automated Self-calibration Imaging Pipeline\footnote{https://science.nrao.edu/facilities/vla/data-processing/pipeline/VIPL}. We used \texttt{CASA}/\texttt{tclean} with custom parameters to image the data. We detect a $\approx 27\sigma$ source, and use the \texttt{pwkit/imtool} program \citep{2017ascl.soft04001W} to measure the flux density of the source, which we find to be $F_{\nu} = 194.0 \pm 7.2 ~\mu$Jy (first reported by \citealt{2023GCN.35114....1S}). A position of $\alpha =~$\ra{22}{09}{33.36} and $\delta =~$\dec{+13}{31}{19.8} is measured, with an uncertainty of $0.2\arcsec$ in each coordinate, which is coincident with the {\it CXO} (Section~\ref{sec:chandra}) and optical (Section~\ref{sec:lris}) positions of the afterglow (Figure~\ref{fig:difference}). 

We initiated a second epoch of multifrequency observations in the S band (1.99--4.01~GHz, central frequency of 3.0~GHz), the C band (3.98--8.02~GHz, central frequency of 6.0~GHz), the X band, and the K$_{u}$ band (12.01--18.16~GHz, central frequency of 15.1~GHz) at $\delta t \approx 3.9~$days, using the same gain and flux calibrators. Unfortunately, owing to severe radio-frequency interference (RFI), the observations experienced extreme gain compression, and those using 3-bit samplers (C, X, and K$_{u}$ bands) are unrecoverable. To recover the S-band 8-bit observations, we applied the gain compression fix in the \texttt{CASA} pipeline, and proceeded with self-calibration and imaging.

\begin{figure}
    \centering
    \includegraphics[width = 0.5\textwidth]{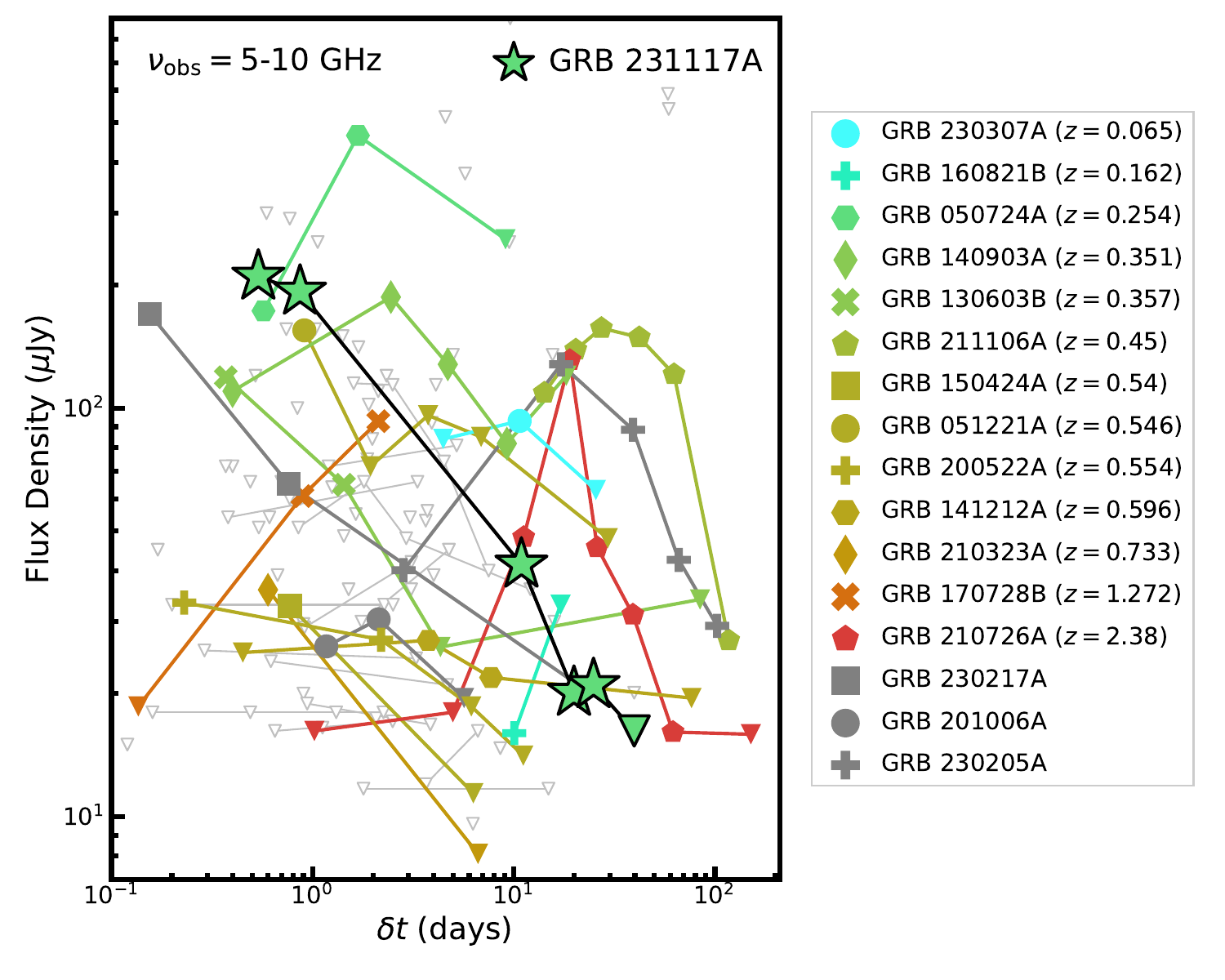}
    \caption{The $10~$GHz VLA light curve of \grb (green stars). Also shown are $5$--$10~$GHz radio afterglows of the 15 other short GRBs with radio detections (colored by redshift) and merger-driven GRB\,230307A. 
    Lines connect observations of the same burst, and triangles represent $3\sigma$ upper limits. Gray triangles represent short GRBs that were observed but not detected at radio wavelengths (also $3\sigma$).}
    \label{fig:RadioAGCompilation}
\end{figure}

We initiated four additional epochs (Programs 23A-296 and 23B-338; PI Schroeder) of multifrequency observations including the S, C, X, and K$_{u}$ bands out to $\sim 39.7~$days. 3C48 for 3C286 were used for flux and bandpass calibration and J2152$+$1734 for complex gain calibration. We thoroughly flag and rerun the pipeline, applying the gain compression fix when applicable. The source faded beyond detection by the final epoch of observations ($\delta t \approx 39.7~$days), confirming the source as the radio afterglow of \grb. We summarize these observations in Table~\ref{tab:RadioAfterglow}, and the $10~$GHz afterglow light curve is plotted in Figure~\ref{fig:RadioAGCompilation}.

\subsubsection{Atacama Large Millimeter/Submillimeter Array}

We observed \grb with the Atacama Large Millimeter/Submillimeter Array (ALMA) at a midtime of $\delta t \approx 1.82~$days (Program 2022.1.00624.T; PI W. Fong). We utilized two 4~GHz wide basebands centered at 91.5 and 103.5 GHz (central frequency of 97.5~GHz), and employed J2207+1652 as the phase and gain calibrator, J2203$+$1725 as the check source, and J2232$+$1143 as the flux and bandpass calibrator. The raw data downloaded from the ALMA archive were calibrated using the ALMA automated pipeline (procedure\_hifa\_cal.xml) in \texttt{CASA}, and imaged using standard techniques. We detect a millimeter source with $F_{\nu} = 111.7\pm 10.2\,\mu$Jy, at $\alpha =~$\ra{22}{09}{33.359}$\pm 0.001$, $\delta =~$\dec{13}{31}{19.77}$\pm 0.02$, coincident with the {\it CXO} (Section~\ref{sec:chandra}), optical afterglow (Section~\ref{sec:lris}), and VLA (Section~\ref{sec:VLA_231117A}) afterglow. We obtained two additional epochs at $\sim 4.5$ and $14.9~$days. The millimeter counterpart is detected in the second epoch but fades beyond detection in the third, confirming it to be the millimeter afterglow. A summary of these observations can be
found in Table~\ref{tab:RadioAfterglow}. With these ALMA observations, \grb\ is the second short GRB to have a mm afterglow detection (after GRB\,211106A; \citealt{2022ApJ...935L..11L}).

\subsubsection{MeerKAT}

We obtained L-band (0.856-1.711 GHz, mean frequency of 1.28 GHz) observations of \grb with the MeerKAT radio telescope (in the Karoo desert, South Africa) at a midtime of $\delta t \approx 9.4~$days through Director's Discretionary time (Program DDT-20231124-SA-01; PIs Rhodes, Schroeder). We used J2232$+$1143 for phase calibration and J1939$-$6342 for flux calibration. Using the pipeline Science Data Processor (SDP) images, we detect a $\sim 9\sigma$ source colocated with the broadband afterglow of $F_{\nu} = 126.1 \pm 14.0\,\mu$Jy, and measure an afterglow position of $\alpha =~$\ra{22}{09}{33.38} and $\delta =~$\dec{+13}{31}{18.9} with a positional uncertainty of $1.2\arcsec$ in each coordinate. We initiated three additional epochs out to $\delta t \approx 40.4~$days, at which time the source faded beyond detection, confirming the source as the afterglow of \grb. A summary of these observations can be found in Table~\ref{tab:RadioAfterglow}. With these data, \grb\ is the second short GRB to have a MeerKAT afterglow detection (after GRB\,210726A; \citealt{2023arXiv230810936S}) and the first to have ALMA, VLA, and MeerKAT detections.

\begin{deluxetable}{cccc}
 \tabletypesize{\footnotesize}
 \tablecolumns{4}
 \tablecaption{GRB 231117A Radio Observations}
 \tablehead{ 
    \colhead{Observatory} &
   \colhead{$\delta t^{\rm{a}}$} &
   \colhead{$\nu^{\rm{b}}$} &
   \colhead{Flux density$^{c}$} \\
      \colhead{} &
   \colhead{(days)} &
   \colhead{(GHz)} &
   \colhead{($\mu$Jy)} 
   }
\startdata 
VLA	 & 0.870 & $10.0$ & $194.0 \pm 7.2$ \\
	 & $3.920$ & $3.0$ & $90.4 \pm 25.1$ \\
	 & $10.91$ & $3.0$ & $62.6 \pm 19.4$ \\
	 &  & $6.0$ & $49.1 \pm 12.5$ \\
	 &  & $10.0$ & $47.6 \pm 7.9$ \\
	 &  & $15.1$ & $51.8 \pm 7.3$ \\
	 & $19.91$ & $3.0$ & $\lesssim 48.6$ \\
	 &  & $6.0$ & $42.8 \pm 5.4$ \\
	 &  & $10.0$ & $18.0 \pm 6.5$ \\
	 &  & $15.1$ & $\lesssim 17.1$ \\
	 & $24.91$ & $6.0$ & $29.1 \pm 6.6$ \\
	 &  & $10.0$ & $20.9 \pm 7.8$ \\
	 & $39.68$ & $6.0$ & $\lesssim 20.1$ \\
	 &  & $10.0$ & $\lesssim 16.8$ \\
\hline
MeerKAT	 & $9.450$ & $1.3$ & $126.1 \pm 14.0$ \\
	 & $15.43$ & $1.3$ & $71.5 \pm 10.3$ \\
	 & $25.41$ & $1.3$ & $31.2 \pm 10.6$ \\
	 & $40.40$ & $1.3$ & $\lesssim 18.3$ \\
\hline
GMRT	 & $31.40$ & $0.55$ & $\lesssim 167.4$ \\
\hline
ALMA	 & $1.820$ & $97.5$ & $116.0 \pm 12.0$ \\
	 & $4.470$ & $97.5$ & $80.3 \pm 9.2$ \\
	 & $14.87$ & $97.5$ & $\lesssim 31.5$ \\
\enddata
\tablecomments{
$^{\rm{a}}$ Midtime of observation compared to {\it Swift}/XRT trigger. \\ $^{\rm{b}}$ Central frequency. \\
$^{c}$ Uncertainties correspond to $1\sigma$ confidence. Upper limits correspond to $3 \sigma$.\\
}
\label{tab:RadioAfterglow}
\end{deluxetable}

\subsubsection{uGMRT}

We obtained one epoch of Band 4 (mean frequency of 550 MHz, bandwidth of 400 MHz) observations of \grb through Director's Discretionary time with the upgraded Giant Meterwave Radio Telescope (uGMRT) at a midtime of $\delta t \approx 31.4~$days (Program ddtC318; PI G. Schroeder). 3C48 was used for bandpass and flux density calibration, and J2250$+$143 for complex gain calibration. We flagged and calibrated the data using standard reduction techniques in \texttt{CASA}, including three rounds of phase-only self-calibration. No source is detected at the afterglow position to a limit of $F_{\nu} \lesssim 167.4~\mu$Jy ($3\sigma$, Table~\ref{tab:RadioAfterglow}).

\subsection{Comparison Sample}
\label{sec:moreradioobs}

\begin{deluxetable*}{cccccccc}
 \tabletypesize{\footnotesize}
 \tablecolumns{7}
 \tablecaption{New VLA Observations of Short GRBs}
 \tablehead{ 
    \colhead{GRB} &
   \colhead{$\delta t^{\rm{a}}$} &
   \colhead{$\nu^{\rm{b}}$} &
   \colhead{Flux density$^{c}$} &
   \colhead{$\alpha$  (error)$^{d}$} &
   \colhead{$\delta$  (error)$^{d}$} &
   \colhead{Location Notes$^{e}$}\\
      \colhead{} &
   \colhead{(days)} &
   \colhead{(GHz)} &
   \colhead{($\mu$Jy)} &
   \colhead{($\arcsec$)} &
   \colhead{($\arcsec$)} &
   \colhead{} &
   }
 \startdata 
\multicolumn{7}{c}{{\it Afterglow Location}}\\
\hline
170127B	 & $1.31$ & $6.05$ & $\lesssim 18.0~(6.0)$ & -- & -- & XRT \\
\hline
170428A	 & $2.07$ & $9.77$ & $\lesssim 17.4~(5.8)$ & -- & -- & Optical$^{1}$ \\
\hline
170524A	 & $0.65$ & $9.77$ & $\lesssim 16.2~(5.4)$ & -- & -- & XRT \\
	 & $2.5$ & $9.77$ & $\lesssim 17.1~(5.7)$ & -- & -- &  \\
\hline
170728B	 & $0.136$ & $6.05$ & $\lesssim 18.6~(6.2)$ & -- & -- & Optical$^{2}$ \\
	 & $0.89$ & $6.05$ & $60.9 \pm 9.5~(6.7)$ & \ra{15}{51}{55.584}~$(2.48)$ & \dec{+70}{07}{21.16}~$(0.47)$ &  \\
	 & $0.89$ & $9.77$ & $126.8 \pm 12.4~(8.7)$ & \ra{15}{51}{55.395}~$(0.91)$ & \dec{+70}{07}{21.33}~$(0.16)$ &  \\
	 & $2.12$ & $6.05$ & $92.9 \pm 8.2~(5.8)$ & \ra{15}{51}{55.461}~$(0.68)$ & \dec{+70}{07}{21.16}~$(0.31)$ &  \\
	 & $2.12$ & $9.77$ & $145.5 \pm 7.0~(4.9)$ & \ra{15}{51}{55.472}~$(0.23)$ & \dec{+70}{07}{21.22}~$(0.11)$ &  \\
\hline
201006A	 & $1.17$ & $6.0$ & $26.1 \pm 5.6~(4.0)$ & \ra{04}{07}{34.426}~$(0.74)$ & \dec{+65}{09}{55.09}~$(0.18)$ & XRT \\
	 & $2.13$ & $6.0$ & $30.4 \pm 5.7~(4.0)$ & \ra{04}{07}{34.472}~$(0.72)$ & \dec{+65}{09}{55.00}~$(0.16)$ &  \\
	 & $5.66$ & $6.0$ & $\lesssim 16.5~(5.5)$ & -- & -- &  \\
\hline
210323A	 & $0.6$ & $6.0$ & $35.9 \pm 8.7~(6.2)$ & \ra{21}{11}{47.240}~$(3.27)$ & \dec{+25}{22}{09.34}~$(2.24)$ & Optical$^{3}$ \\
	 & $6.64$ & $6.0$ & $\lesssim 8.1~(2.7)$ & -- & -- &  \\
\hline
210919A	 & $8.57$ & $5.92$ & $\lesssim 14.7~(4.9)$ & -- & -- & Optical$^{4}$ \\
\hline
211023B	 & $3.65$ & $6.0$ & $\lesssim 12.0~(4.0)$ & -- & -- & Optical$^{5}$ \\
	 & $6.63$ & $6.0$ & $\lesssim 16.2~(5.4)$ & -- & -- &  \\
\hline
230205A	 & $2.83$ & $6.0$ & $40.1 \pm 8.8~(6.2)$ & \ra{13}{28}{16.843}~$(0.50)$ & \dec{+46}{43}{32.85}~$(0.22)$ & Radio$^{6}$ \\
	 & $17.19$ & $6.0$ & $128.0 \pm 6.3~(4.4)$ & \ra{13}{28}{16.836}~$(0.07)$ & \dec{+46}{43}{33.12}~$(0.05)$ &  \\
	 & $18.88$ & $10.0$ & $139.8 \pm 8.3~(5.9)$ & \ra{13}{28}{16.838}~$(0.05)$ & \dec{+46}{43}{33.11}~$(0.03)$ &  \\
	 & $39.16$ & $6.0$ & $88.5 \pm 9.4~(6.6)$ & \ra{13}{28}{16.831}~$(0.25)$ & \dec{+46}{43}{33.17}~$(0.10)$ &  \\
	 & $66.08$ & $6.0$ & $42.5 \pm 5.4~(3.8)$ & \ra{13}{28}{16.830}~$(0.28)$ & \dec{+46}{43}{33.13}~$(0.11)$ &  \\
	 & $102.57$ & $6.0$ & $29.3 \pm 7.7~(5.4)$ & \ra{13}{28}{16.862}~$(0.55)$ & \dec{+46}{43}{33.47}~$(0.21)$ &  \\
\hline
230217A	 & $0.78$ & $6.0$ & $65.2 \pm 6.9~(4.9)$ & \ra{18}{43}{04.947}~$(0.23)$ & \dec{-28}{50}{16.56}~$(0.11)$ & Optical$^{7}$ \\
	 & $21.74$ & $6.0$ & $\lesssim 20.7~(6.9)$ & -- & -- &  \\
\hline
230228A	 & $1.79$ & $6.0$ & $\lesssim 11.7~(3.9)$ & -- & -- & XRT \\
	 & $14.87$ & $6.0$ & $\lesssim 11.7~(3.9)$ & -- & -- &  \\
\hline
\hline
\multicolumn{7}{c}{{\it Other Radio Sources}}\\
\hline
\hline
170325A	 & $0.94$ & $6.05$ & $46.3 \pm 7.1~(5.0)$ & \ra{08}{29}{59.997}~$(1.82)$ & \dec{+20}{32}{06.55}~$(1.48)$ & Radio Source 1 \\
	 & $3.85$ & $6.05$ & $\lesssim 23.2~(7.0)$ & -- & -- &  \\
\hline
	 & $0.94$ & $6.05$ & $60.7 \pm 8.6~(6.0)$ & \ra{08}{29}{54.713}~$(1.69)$ & \dec{+20}{31}{27.68}~$(1.37)$ & Radio Source 2 \\
	 & $3.85$ & $6.05$ & $57.7 \pm 11.2~(7.9)$ & \ra{08}{29}{54.697}~$(1.97)$ & \dec{+20}{31}{26.59}~$(1.76)$ &  \\
\hline
	 & $0.94$ & $6.05$ & $38.5 \pm 6.4~(4.5)$ & \ra{08}{29}{54.381}~$(2.01)$ & \dec{+20}{32}{16.93}~$(1.63)$ & Radio Source 3 \\
	 & $3.85$ & $6.05$ & $29.4 \pm 10.2~(7.2)$ & \ra{08}{29}{54.257}~$(3.52)$ & \dec{+20}{32}{14.49}~$(3.14)$ &  \\
\hline
	 & $0.94$ & $6.05$ & $\lesssim 18.9~(6.3)$ & -- & -- & Optical?$^{8}$ \\
	 & $3.85$ & $6.05$ & $\lesssim 16.8~(5.6)$ & -- & -- &  \\
\hline
\hline
201006A	 & $1.17$ & $6.0$ & $\lesssim 17.1~(5.7)$ & -- & -- & Coherent Radio$^{9}$ \\
	 & $2.13$ & $6.0$ & $21.7 \pm 5.9~(4.2)$ & \ra{04}{07}{36.661}~$(1.04)$ & \dec{+65}{09}{29.45}~$(0.23)$ &  \\
	 & $5.66$ & $6.0$ & $\lesssim 19.5~(6.5)$ & -- & -- &  \\
\hline
\hline
210919A	 & $8.57$ & $5.92$ & $38.9 \pm 10.8~(5.5)$ & \ra{05}{21}{01.943}~$(0.42)$ & \dec{+01}{18}{40.22}~$(0.41)$ & Host Galaxy$^{10}$ \\
 \enddata
\tablecomments{
$^{\rm{a}}$ Midtime of entire observation compared to {\it Swift}/BAT trigger. \\ $^{\rm{b}}$ Central frequency. \\
$^{c}$ Uncertainties correspond to $1\sigma$ confidence. Upper limits correspond to $3 \sigma$. The RMS of the image is in parentheses. \\
$^{d}$ In the case where a radio source is detected, the right ascension ($\alpha$) and declination ($\delta$) of the source are listed, with the positional uncertainty in arcsec listed in parentheses. \\
$^{e}$ Source position notes and references. All XRT positions are from \citet{2009MNRAS.397.1177E}.  
$^{1}$\citet{2017GCN.21050....1B}, 
$^{2}$\citet{2017GCN.21373....1D}, 
$^{3}$\citet{2021GCN.29703....1M}, 
$^{4}$\citet{2021GCN.30883....1K}, 
$^{5}$\citet{2021GCN.30960....1D}, 
$^{6}$\citet{2023GCN.33309....1S}, 
$^{7}$\citet{2023GCN.33374....1D},
$^{8}$\citet{2017GCN.20978....1M},
$^{9}$\citet{2023arXiv231204237R},
$^{10}$\citet{2022ApJ...940...56F}. \\
}
\label{tab:otherradio_data}
\end{deluxetable*}

As part of a comparison sample to GRB\,231117A, we present all unpublished VLA observations\footnote{We remove GRB\,191031D owing to complications with the data on account of the observations being taken during VLA configuration changes.} of short (or possibly short) GRBs from our programs. This comprises 12 additional events, obtained between 2017 and 2023 under programs 17A-218, 19B-217, 21B-198, 20B-057 (PI W. Fong), and 23A-296 (PI G. Schroeder). We download the \texttt{CASA} Pipeline Products and image the data using custom routines with \texttt{CASA}/\texttt{tclean}. A summary of the observations is given in Table~\ref{tab:otherradio_data}. In addition to three previously known detections \citep[GRBs\,170728B, 230205A, and 230217A;][]{2017GCN.21395....1F, 2023GCN.33372....1S, 2023GCN.33358....1S}, there are low-significance possible radio afterglow detections of GRBs\,201006A and GRB\,210323A, which are detected at $\lesssim 5\sigma$ confidence in one or more epochs, bringing the total radio-detected short GRB sample to 16 bursts (Figure~\ref{fig:RadioAGCompilation}). We also present new deep upper limits for 6 bursts, and inconclusive radio observations for 1 burst which was not well localized. We briefly summarize events with interesting results below. 

{\it GRB\,170325A:} GRB\,170325A did not trigger {\it Swift} owing to the spacecraft slewing at the time of the burst, and was instead discovered through ground analysis of {\it Swift}/BAT data \citep{2017GCN.20938....1P}. As a result, there were no XRT observations of  GRB\,170325A,
but a candidate optical counterpart which faded marginally was found within the BAT localization \citep{2017GCN.20978....1M}. We observed the position of GRB\,170325A at a mean frequency of 6.05~GHz for two epochs ($\delta t \approx 0.9, 3.9~$days), and the VLA field of view covered the entire BAT localization. In our first epoch of observations, we detect three radio sources at $\gtrsim 5 \sigma$. For one of the sources (``Radio Source 1'') we measure a constant flux density across both epochs, and we therefore conclude that the radio source is unrelated to the GRB\footnote{This source is also coincident with point source SDSS J082954.54+203133.1.}. The other two sources (``Radio Source 2'' and ``Radio Source 3'') fade significantly by the second epoch, and thus we cannot conclude whether either source is related to the GRB. At the location of the candidate optical counterpart, we do not detect any source. As we are unable to confidently associate any of the radio sources with GRB\,170325A, we proceed without this burst in subsequent analysis. We provide the flux-density measurements for the three radio sources and the candidate optical counterpart in Table~\ref{tab:otherradio_data}.

{\it GRB\,201006A:} We observed the position of GRB\,201006A for three epochs ($\delta t \approx 1.2, 2.1, 5.7~$days) at a mean frequency of 6.0 GHz. In the first two epochs, we detect a radio source ($\sim 5\sigma$) just outside the edge of the the XRT localization region ($\sim 2.8''$ offset from XRT center with 90\% containment corresponding to a $2.1''$ radius; \citealt{2009MNRAS.397.1177E}). The radio source is not detected in the third epoch despite similarly deep observations. We therefore claim this source as the radio afterglow of GRB\,201006A. We also detect a radio source ($\sim 4 \sigma$) at $\alpha =~$\ra{04}{07}{36.661} $\pm 1.04\arcsec$, $\delta =~$\dec{+65}{09}{29.45}$ \pm 0.23\arcsec$, which is within the $\sim 5\arcsec$ localization of the coherent radio source detected by the Low-Frequency Array \citep[LOFAR;][]{2023arXiv231204237R} in the second epoch. This source also appears to fade by the third epoch. However, we note that the $3\sigma$ root-mean square (RMS) in the third epoch is comparable to the peak flux we measure in the second epoch, so there is only marginal evidence for fading in our data. We present flux-density measurements and positions of both sources in Table~\ref{tab:otherradio_data}. 

{\it GRB\,210323A:} We observed the position of GRB\,210323A for two epochs ($\delta t \approx 0.6, 6.6~$days) at a mean frequency of 6.0~GHz. In the first epoch, we detect a radio source ($\sim 4 \sigma$) coincident with the location of the optical and X-ray afterglow \citep{2021GCN.29703....1M}. We do not detect this source to deep limits in the second epoch, whereas static sources of similar brightness are detected in both epochs. We therefore claim this source as the radio afterglow of GRB\,210323A, and present flux-density measurements and positions in Table~\ref{tab:otherradio_data}.

{\it GRB\,210919A:} We observed the position of GRB\,210919A at $\delta t \approx 8.6~$days at a mean frequency of 5.92~GHz. We do not detect any radio counterpart at the location of the optical afterglow \citep{2021GCN.30883....1K}. However, we detect a radio source ($\sim 4 \sigma$) coincident with the host galaxy of GRB\,210919A \citep{2022ApJ...940...56F}, finding a flux density of $F_{\nu} = 38.9 \pm 10.8 \mu$Jy\footnote{This radio source appears moderately extended, and a Gaussian shape is preferred by \texttt{imtool}.}. Given the large offset ($\sim 13''$) from the optical afterglow position \citep{2021GCN.30883....1K}, this source is unrelated to the afterglow of GRB\,210919A. If we assume the radio emission is due to star formation within the host galaxy, this results in a radio star-formation rate (SFR; \citealt{2016A&A...593A..17G}) of $11.0 \pm 3.4\,M_\odot$~yr$^{-1}$ ($z = 0.2415$), $\sim 40$ times higher than the optically derived SFR of $0.3 \pm 0.03\,M_\odot$~yr$^{-1}$ \citep{2022ApJ...940...57N}.

In addition to the GRBs mentioned above, we also consider all published radio observations of short GRBs, which includes radio detections of 16 short GRBs and 1 merger-driven GRB \citep[][this work]{2005Natur.438..988B, 2006ApJ...650..261S, 2014ApJ...780..118F, 2015ApJ...815..102F, 2017GCN.21395....1F, 2019ApJ...883...48L, 2021ApJ...906..127F, 2022ApJ...935L..11L, 2023GCN.33372....1S, 2023GCN.33358....1S, 
2023arXiv230810936S, 2023GCN.35097....1R}.

\section{Host Galaxy}
\label{sec:HostGalaxy_231117A}

\subsection{Galactocentric Offset}
\label{sec:host-offset}

To characterize the precise location of GRB\,231117A relative to its host galaxy, we measure the projected offset between the centroid of its host galaxy and optical afterglow within our Keck/LRIS $G$-band imaging.  Optical data provide the best constraint on this quantity for several reasons: they can be aligned to a common absolute astrometric frame \citep[in this case, the {\it Gaia} DR3 frame;][]{2023A&A...674A...1G}, our difference imaging allows us to precisely centroid our optical detection of the afterglow, and our final template image allows us to measure the position of the host galaxy itself.
We use our $I$-band template image, which we assume has no afterglow flux, to centroid on the position of the host galaxy using the  methods detailed in Section~\ref{sec:lris}.  This results in a host-galaxy position of $\alpha =~$\ra{22}{09}{33.3370} $\pm 0.0016$ and $\delta =~$\dec{+13}{31}{19.469} $\pm 0.015$ (J2000).  
Thus, our combined nominal offset of the afterglow from the host galaxy using optical data is $\Delta \alpha = +0.268 \pm 0.013\arcsec$ and $\Delta\delta = +0.340 \pm 0.015\arcsec$.

\subsection{Host-Galaxy Properties}
\label{sec:prospector}

\begin{figure*}
    \centering
    \includegraphics[width = 0.9\textwidth]{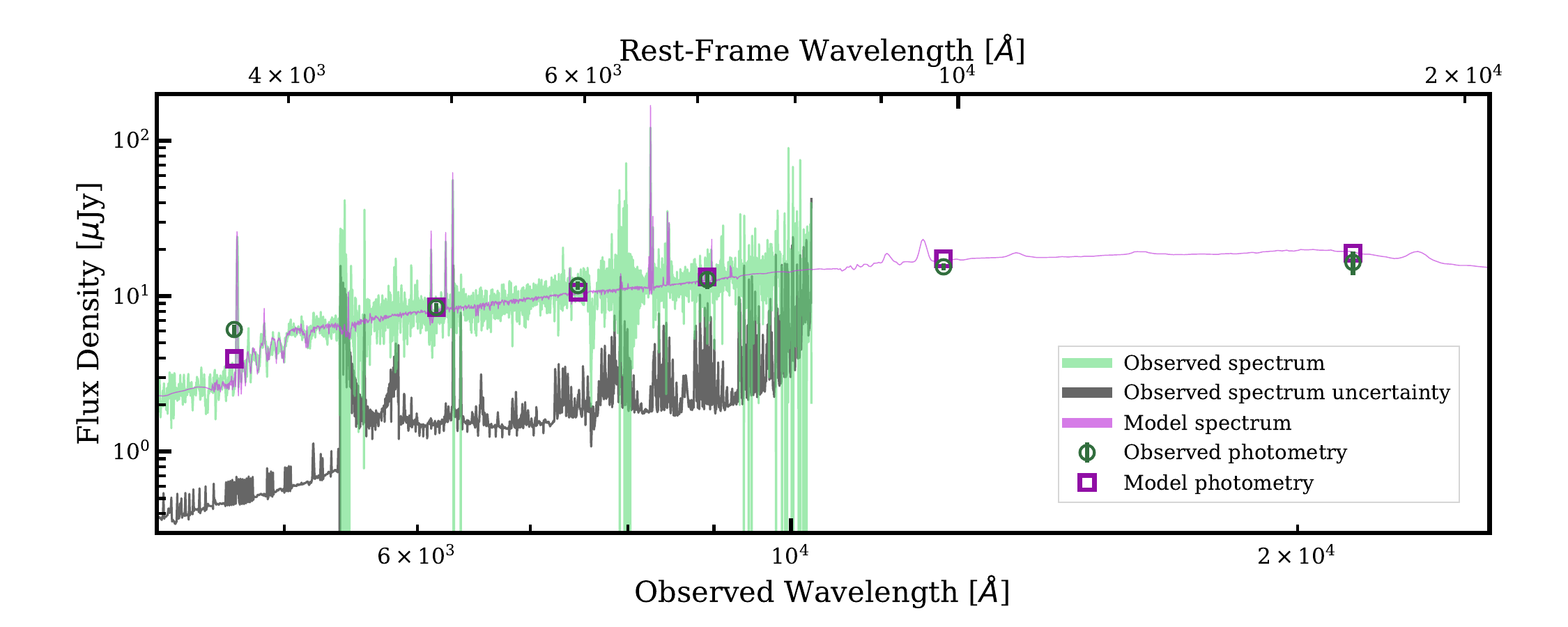}
    \caption{The observed $grizJK$ photometry (green data points), spectrum (green line), and error spectrum (gray line) of the host galaxy of \grb, compared to the \texttt{Prospector}-produced model spectrum and photometry (purple line and squares, respectively). }
    \label{fig:Prospector}
\end{figure*}

We model the stellar population properties of the host galaxy of \grb\ with the host-galaxy photometry (Table~\ref{tab:optical}) and the Keck/LRIS host spectrum (Section~\ref{sec:lris}) as inputs, using the stellar population inference code \texttt{Prospector} \citep{2019ApJ...877..140L, 2021ApJS..254...22J}. We apply a \texttt{dynesty} nested sampling fitting routine \citep{2020MNRAS.493.3132S} within  \texttt{Prospector} to determine the posterior distributions on the stellar population properties of interest. Internally, \texttt{Prospector} employs \texttt{FSPS} and \texttt{python-FSPS} to generate model spectral energy distributions \citep[SEDs;[]{2009ApJ...699..486C, 2010ApJ...712..833C}. We assume a Milky Way extinction law \citep{1989ApJ...345..245C}, a Chabrier initial mass function (IMF; \citealt{2003PASP..115..763C}), and a parametric delayed-$\tau$ star-formation history (SFH). The spectral continuum is modeled with a tenth-order Chebyshev polynomial and the spectral line strengths are constrained through a nebular emission model. We refer the reader to \citet{2022ApJ...940...57N} for further details on the stellar population modeling methods. The resulting SED fit is presented in Figure~\ref{fig:Prospector}. 

We determine median and 68\% confidence intervals for the host's stellar mass ($M_*$), stellar population age ($t_m$), star-formation rate (SFR), stellar ($Z_*$) and gas-phase ($Z_\textrm{gas}$) metallicities, and dust extinction (reported as a $V$-band magnitude, $A_V$). We find $\log(M_*/M_\odot) = 9.16^{+0.05}_{-0.05}$, $t_m = 1.62^{+0.34}_{-0.27}$~Gyr, SFR $= 0.39^{+0.04}_{-0.03} \,M_\odot$~yr$^{-1}$,  $\log(Z_*/Z_\odot) = -0.97^{+0.04}_{-0.02}$, $\log(Z_\textrm{gas}/Z_\odot) = -0.33^{+0.02}_{-0.03}$, and $A_V = 0.82^{+0.09}_{-0.09}$~mag. Note that these are consistent with the median and 68\% confidence intervals for the stellar population properties of the full short GRB population \citep{2022ApJ...940...57N}. Thus, the host galaxy of \grb is a fairly typical short GRB host, and the \texttt{Prospector}-derived stellar population properties confirm the star-forming classification of the galaxy \citep{2022ApJ...926..134T, 2022ApJ...940...57N}.

\subsection{Afterglow Spectral Subtraction}\label{sec:ag-subtraction}

We now use the second LRIS epoch at $\sim 48.4~$days, dominated by host-galaxy light, to estimate and subtract the host contribution from the first epoch at $\sim 1.1~$days and obtain an ``afterglow-only'' spectrum over the range $\sim 3000$--10,000~\AA\ at $\sim 1.1~$days. 
This requires that we properly scale the relative host-galaxy flux between both spectra, whose overall flux calibration can vary significantly owing to placement of the slit, slit losses due to seeing, and the extraction procedure used in the reductions described in Section~\ref{sec:lris}.

We scale the host-galaxy flux between the two epochs by identifying nebular emission lines in both epochs and calculating the flux for each line in both spectra, specifically for [O~\textsc{ii}] $\lambda\lambda$3726, 3929, H$\beta$, H$\alpha$, and [O~\textsc{iii}] $\lambda$5007.  At the redshift of \grb, these lines cover $\sim 4685$--9119~\AA, spanning most of the wavelength range of our spectra. Under the assumption that they arise solely in the host galaxy and should not vary across our two epochs, this enables us to subtract off cleanly the host-galaxy emission-line and continuum flux in the first epoch.  We further adopt a wavelength-dependent scale factor that is fitted to the ratio between the nebular emission-line fluxes as a function of wavelength using a 1D polynomial.  This polynomial accounts for small differences in the flux calibration or sources of extinction such as the slightly different airmasses during the two observations. We correct the afterglow spectrum for Galactic dust \citep{1989ApJ...345..245C} using $A_V = 0.196~$mag \citep{2011ApJ...737..103S}. The final, subtracted afterglow spectrum is shown in Fig.~\ref{fig:afterglow}, where 
some of the strongest emission lines are identified. The afterglow spectrum at $\delta t \approx 1.1~$days can be fit with a single power law, resulting in a spectral index of $\beta_O = -0.86 \pm 0.01$. This slope is shown in Fig.~\ref{fig:afterglow}, for comparison with the subtracted observations.

\section{Afterglow Modeling}
\label{sec:AGModeling}

We interpret the radio through X-ray afterglow of \grb in the context of synchrotron emission from a relativistic forward shock (FS) produced by the GRB jet interacting with the surrounding environment \citep[e.g.,][]{1998ApJ...497L..17S, 1999ApJ...523..177W, 2000ApJ...543...66P, 2002ApJ...568..820G}. The FS accelerates the electrons into a nonthermal power-law distribution ($N(\gamma_e) \propto \gamma_e^{-p}$, where $p$ is the power-law index). The emission can be described by the peak flux ($F_{\nu, \rm m}$) as well as three break frequencies: the self-absorption frequency ($\nu_{\rm sa}$), the peak frequency ($\nu_{\rm m}$), and the cooling frequency ($\nu_{\rm c}$). The values of these break frequencies and peak flux are set by the isotropic equivalent kinetic energy of the burst ($E_{\rm K, iso}$), the circumburst density of the GRB ($\rho = A r^{-k}$; in this case we assume a constant-density environment where $k = 0$ and $A = m_p n_0$), $p$, and the microphysical parameters that describe the fraction of energy imparted on the electrons ($\epsilon_{\rm e}$) and magnetic field ($\epsilon_{\rm B}$). Throughout this analysis, we use the convention $F_{\nu} \propto t^{\alpha} \nu^{\beta}$.

While the standard FS model described above assumes a spherically symmetric blast wave, we additionally consider the effects of collimation on the afterglow behavior. When a GRB jet decelerates sufficiently such that the angular size of the beaming angle ($\theta_{\rm beam} = 1/\Gamma$, where $\Gamma$ is the bulk Lorentz factor of the post-shock fluid) approaches the size of the true opening angle of the jet ($\theta_{\rm j}$), an achromatic break is observed in the afterglow light curve. This ``jet break'' occurs at time $t_{\rm jet}$, and given $E_{\rm K, iso}$, $n_0$, and $t_{\rm jet}$, we can calculate $\theta_{\rm j}$ \citep{1999ApJ...525..737R, 1999ApJ...519L..17S}. After the jet break, the light curves at observing frequencies $\nu_{\rm obs} > \nu_{\rm m}$ are expected to follow $F_{\nu} \propto t^{-p}$.
Typically, at the beginning of the detected afterglow, we expect $\nu_{\rm m}$ to be between the radio and optical bands, evolving through the radio to lower frequencies at later times. 

\subsection{Basic Considerations}
\label{sec:BasicConsiderations}

The X-ray light curve exhibits a flare and a plateau stage ($\delta t \approx 0.05$--$0.8$~days). This behavior is not consistent with a standard FS afterglow, and could indicate an alternative dominant emission mechanism at early times \citep[e.g.,][]{2006ApJ...642..354Z, 2013MNRAS.428..729M, 2013MNRAS.430.1061R, 2015MNRAS.448..629G}. Therefore, for this analysis and our initial afterglow fitting, we ignore all data at $\delta t \lesssim 1~$day. 

\begin{figure*}
    \centering
    \includegraphics[width = 0.9\textwidth]{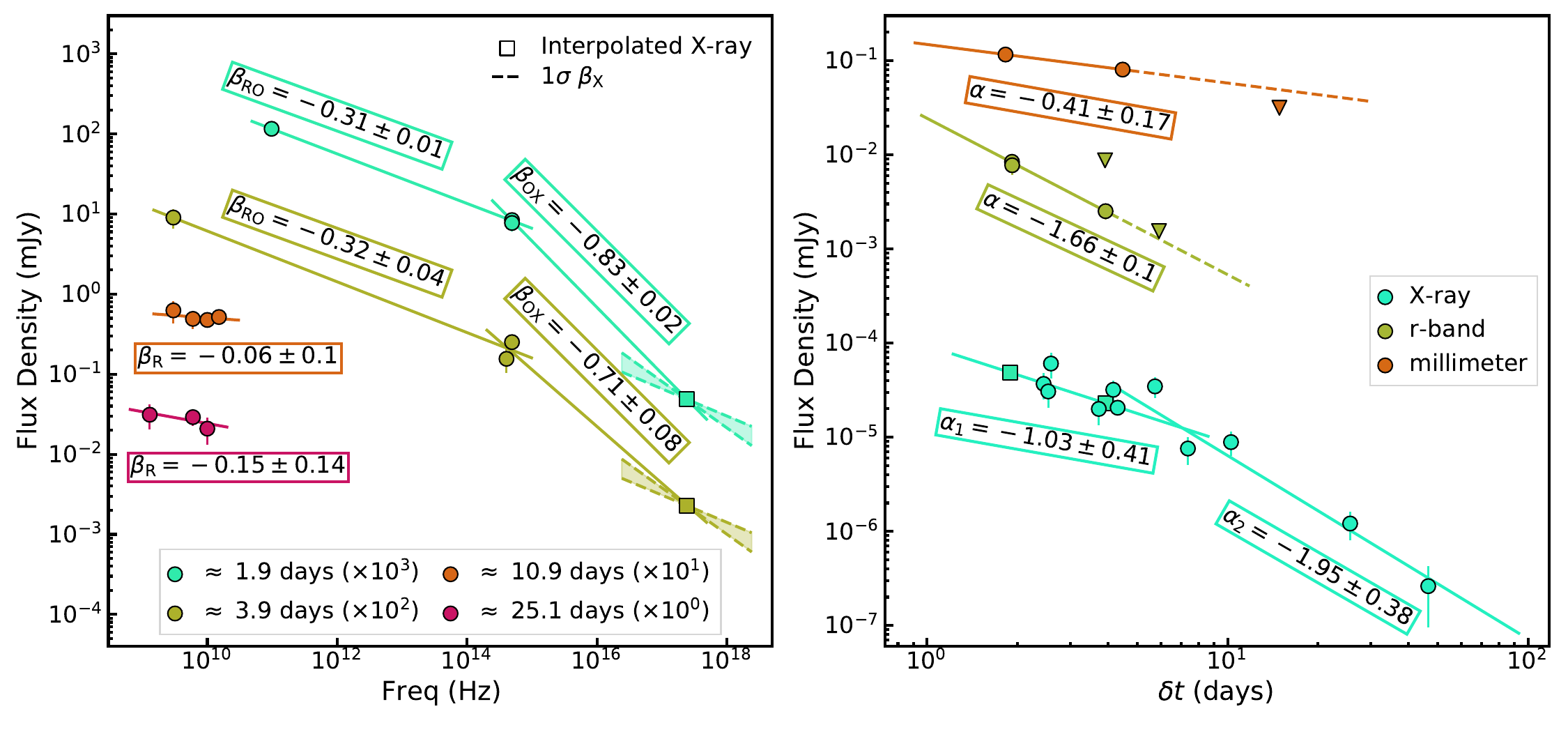}
    \caption{{\emph Left:} SEDs for nearly simultaneous ($\Delta t/t \lesssim 0.3$) observations across our radio, optical, and X-ray data. The dashed line and shaded region represent the $1\sigma$ uncertainty in the measurement of $\beta_{\rm X}$ (Section~\ref{sec:chandra}). Solid lines show power-law fits between the radio ($\beta_{\rm R}$), radio-to-optical ($\beta_{\rm RO}$), and optical-to-X-ray ($\beta_{\rm OX}$) data.
    \\
    {\emph Right:} Light curves of our millimeter, optical ($r$ band), and X-ray afterglow at $\delta t \gtrsim 1~$days. Circles represent detections, squares are interpolated afterglow flux densities, and triangles indicate nondetections ($3\sigma$). Solid lines represent power-law fits to the detected afterglow, while dashed lines provide an extrapolation past the last detection. }
    \label{fig:Spectra_LC_Fits}
\end{figure*}

We first explore the radio (R), millimeter (mm), and optical (O) afterglows to determine the location of $\nu_{\rm m}$. If $\nu_{\rm m} < \nu_{\rm R} < \nu_{\rm O}$, we would expect the radio-to-optical spectral index to  be $\beta_{\rm RO} \lesssim -0.5$ ($\beta_{\rm RO} = (1-p)/2$; here we assume $p \gtrsim 2$; \citealt{2002ApJ...568..820G}), whereas if $\nu_{\rm R} <\nu_{\rm m} < \nu_{\rm O}$, we would expect $\beta_{\rm RO} \gtrsim -0.5$. In Figure~\ref{fig:Spectra_LC_Fits} (left), we construct SEDs at common observation times of $\delta t \approx 1.9~$days and $\delta t \approx 3.9~$days, and measure $\beta_{\rm RO} \approx -0.31$ and $\beta_{\rm RO} \approx -0.32$, respectively. This indicates $\nu_{\rm R} < \nu_{\rm m} < \nu_{\rm O}$, and the observed radio and optical fluxes together imply $\nu_{\rm m}\approx 2 \times 10^{12}~$Hz with a peak flux density of $F_{\nu,\rm m}\approx 310~\mu$Jy at $\delta  t \approx 1.9~$days. 

We can also constrain the evolution of $\nu_{\rm m}$ to later times with radio observations (at $\delta t \approx 10.9,~25.2$~days). As $\nu_{\rm m}$ evolves to lower frequencies and through the radio band, $\beta_{\rm R}$ is expected to change from $\sim 1/3$ ($\nu_{\rm R} < \nu_{\rm m}$) to $\lesssim -0.5$ \citep[$\nu_{\rm m} < \nu_{\rm R}$ for $p \gtrsim 2$;][]{2002ApJ...568..820G}. Our radio SED at $\delta t \approx 10.9~$days, spanning $\nu_{\rm obs} \approx 3$--$15~$GHz, exhibits $\beta_{\rm R} \approx -0.06$. Similarly, at $\delta t \approx 25.2~$days ($\nu_{\rm obs} \approx 1.3$--$10~$GHz), we measure $\beta_{\rm R} \approx -0.15$  (Figure~\ref{fig:Spectra_LC_Fits}). Thus, the change in $\beta_{\rm R}$ to lower values indicates that $\nu_{\rm m}$ is evolving to lower frequencies and passing through the radio band at $\delta t \approx 10.9~$days. Additionally, the SED at $\delta t \approx 10.9~$days indicates $F_{\nu, \rm m} \approx 50~\mu$Jy. Prior to a jet break, $F_{\nu, \rm m}$ is expected to remain constant \citep{2002ApJ...568..820G}, whereas afterward $F_{\nu, \rm m} \propto t^{-1}$ \citep{1999ApJ...519L..17S}. Comparing our predicted values of $F_{\nu, \rm m}$ at $\sim 1.9~$days ($\sim 310~\mu$Jy) and $\sim 10.9~$days ($\sim 50~\mu$Jy), we find $F_{\nu, \rm m}$ evolves with a temporal index $\sim -1$, and therefore we expect to find evidence of a jet break in our afterglow light curves at $\delta t \gtrsim 1.9~$days.

Extrapolating from $F_{\nu,\rm m}$ (which remains constant in time prior to a jet break, \citealt{2002ApJ...568..820G}) and $\nu_{\rm m}$ ($\propto t^{-3/2}$), the expected $10~$GHz flux density at the time of our first VLA detection ($\delta t \approx 0.9~$days) is $\sim 40~\mu$Jy ($F_{\nu} \propto \nu^{1/3}$, when $\nu_{\rm sa} < \nu_{\rm obs} < \nu_{\rm m}$, \citealt{2002ApJ...568..820G}), $\sim 5$ times lower than our measured value of $\sim 190~\mu$Jy. Therefore, similar to the X-rays, we do not expect the standard FS to be able to account for the $10~$GHz afterglow at $\delta t \lesssim 1~$days.

We next explore the mm, optical, and X-ray (X) afterglow to search for further evidence of a jet break. The expected temporal index for $\nu_{\rm sa} < \nu_{\rm obs}< \nu_{\rm m}$ prior to a jet break is $\alpha = 0.5$ \citep{2002ApJ...568..820G}, whereas $\alpha = -1/3$ \citep{1999ApJ...519L..17S} after a jet break. We measure $\alpha_{\rm mm} \approx -0.41$ at $\delta t \approx 1.8$--$4.5~$days (Figure~\ref{fig:Spectra_LC_Fits}; right), similar to the expected value post jet break, indicating $t_{\rm jet} \lesssim 4.5~$days. The mm data require a steepening by the third observation at $\delta t \approx 14.7~$days, indicating $\nu_{\rm m} < \nu_{\rm mm}$ by $\delta t \approx 14.7~$days, consistent with our expectations from the radio SEDs at $\delta t \approx 10.9, 25.1~$days. 
The X-ray afterglow at $\delta t \gtrsim 4.5~$days can be fit with a single power law with an index of $\alpha_{\rm X} = -1.95 \pm 0.38$ (Figure~\ref{fig:Spectra_LC_Fits}), indicating $p = 1.95 \pm 0.38$. The optical band also exhibits a steep decay with $\alpha_{\rm O} \approx -1.68$ starting at $\delta t \approx 1.9$~days, indicative of a jet break, or transition to a jet break, around this time. 

With an estimate of $p$, we can utilize the optical and X-ray SEDs to determine the location of $\nu_{\rm c}$. 
If the optical and X-ray afterglows are in the same spectral regime ($\nu_{\rm O} < \nu_{\rm X} < \nu_{\rm c}$), the optical-to-X-ray spectral index is $\beta_{\rm OX} \approx -0.48$ ($\beta_{\rm OX} = (1-p)/2$ for $p \approx 1.95$). However, if they are in different spectral regimes ($\nu_{\rm O} < \nu_{\rm c} < \nu_{\rm X}$), we would expect $-0.98 \lesssim \beta_{\rm OX} \lesssim -0.48$ ($\beta_{\rm X} = -p/2$ for $p \approx 1.95$; \citealt{2002ApJ...568..820G}). We construct SEDs at $\delta t \approx 1.9~$days and $\delta t \approx 3.9~$days, the time of our optical detections, and interpolate the X-ray data to these times assuming the X-ray afterglow follows a single power-law decay at $1 \lesssim \delta t \lesssim 4.5~$days with temporal index $\alpha_{\rm X} \approx -1.03$. We find $\beta_{\rm OX} \approx -0.83$ at $\delta t \approx 1.9~$days and $\beta_{\rm OX} \approx -0.71$ at $\delta t \approx 3.9~$days (Figure~\ref{fig:Spectra_LC_Fits}), consistent with $ \nu_{\rm O} < \nu_{\rm c} < \nu_{\rm X}$. The measured value of $\beta_{\rm X} = -0.46 \pm 0.12$ (Section~\ref{sec:chandra}) is inconsistent with our expected value of $\sim -0.98$. This can be reconciled if $\nu_{\rm X} \approx \nu_{\rm c}$, which would lead to an intermediate value of $\beta_{\rm X}$ between $\sim -0.48$ and $\sim -0.98$, consistent with our measured values. We therefore conclude that $ \nu_{\rm O} < \nu_{\rm c} \approx \nu_{\rm X}$.
As previously mentioned, the standard FS model is likely not the primary source of emission at early times, so it is unsurprising that the optical slope ($\beta_{\rm O} = -0.86 \pm 0.01$) is steeper than the expected value ($\sim -0.48$).
We return to this point in Section~\ref{sec:EnjRS}.

In summary, from our analysis of the radio, millimeter, optical, and X-ray afterglows at $\delta t \gtrsim 1~$days, we expect $\nu_{\rm R} < \nu_{\rm mm} < \nu_{\rm m} < \nu_{\rm O} < \nu_{\rm c} \approx \nu_{\rm X}$ at $\delta t \lesssim 4~$days. Additionally, we expect $p = 1.95 \pm 0.38$ and $1.9~{\rm days} \lesssim t_{\rm jet} \lesssim 4.5~{\rm days}$. Furthermore, given that the X-ray light curve exhibits a plateau and the $9.0$--$10.0~$GHz observations are in excess of the expected flux from the FS at $\delta t \lesssim 1~$days, we do not expect the broadband afterglow at $\delta t \lesssim 1~$days to be well fit with a standard FS model.

\begin{deluxetable}{ccc}
 \tabletypesize{\footnotesize}
 \tablecolumns{3}
 \tablecaption{Forward-Shock Parameters}
 \tablehead{  
    % \colhead{} &
    % \twocolhead{Redshift Fixed}% & 
    % \colhead{} &
    % \twocolhead{Redshift Free}
    % % \colhead{} 
    \colhead{Parameter} &
    \colhead{Best-Fit Model} &
    \colhead{MCMC Results} %&
 %   % \colhead{Best Fit Model} &
 %   % \colhead{MCMC Results}
 %   %\\
   }
 \startdata 
 % Parameter & Best Fit Model & MCMC Results & Best Fit Model & MCMC Results \\
 % \hline
 % $z$ & $2.38$ & -- & $2.74$ & $2.29^{+0.44}_{-0.63}$\\
$p$ & $2.29$ & $2.24^{+0.07}_{-0.08}$ \\
$E_{\rm K, iso}~(10^{52}~{\rm erg})$ & $2.7\times 10^{-1}$ &$1.5^{+1.3}_{-0.7} \times 10^{-1}$ \\
$n_0~({\rm cm}^{-3})$ & $7.3\times 10^{-2}$ &$4.8^{+1.8}_{-2.2} \times 10^{-2}$ \\
$\epsilon_{\rm e}$ & $9.7\times 10^{-1}$ &$8.5^{+1.0}_{-1.4} \times 10^{-1}$ \\
$\epsilon_{\rm B}$ & $7.2\times 10^{-4}$ &$32.1^{+140.2}_{-24.3} \times 10^{-4}$ \\
$t_{\rm jet}~({\rm days})$ & $2.4$ & $2.6^{+0.4}_{-0.4}$ \\
$\theta_{\rm jet}~({\rm deg})$ & $10.4$ & $10.6^{+0.9}_{-0.9}$ \\
$E_{\rm K}~(10^{52}~{\rm erg})$ & $4.4\times 10^{-3}$ &$2.6^{+1.8}_{-1.1} \times 10^{-3}$ \\
$A_{V, \rm GRB}~ ($\rm mag$)$ & $0$ & $0$ \\
\hline
\multicolumn{3}{c}{Break Frequencies and Peak Flux at $\delta t = 1$~days} \\
\hline
$\nu_{\rm sa}~$(Hz) & $5.1 \times 10^{7}$ & -- \\
$\nu_{\rm m}~$(Hz) & $4.3 \times 10^{12}$ & -- \\
$\nu_{\rm c}~$(Hz) & $1.8 \times 10^{17}$ & -- \\
$F_{\nu, \rm m}~$($\mu$Jy) & $350$ & -- \\
\enddata
\tablecomments{\emph{Top:} The best-fit
and summary statistics (median and 68\% credible intervals) parameters from the marginalized posterior density functions of the  FS afterglow parameters from our MCMC modeling. The parameters of the best-fit model may differ from the summary statistics as the former is the peak of the likelihood distribution and the latter is calculated from the full marginalized posterior density functions of each parameter.
\emph{Bottom:} The break frequencies and peak flux of the FS from the best-fit parameters at $\delta t = 1~$days}
\label{tab:FS_fit}
\end{deluxetable}

\subsection{MCMC Modeling}
\label{sec:MCMC}

\begin{figure}
    \centering
    \includegraphics[width = 0.5\textwidth]{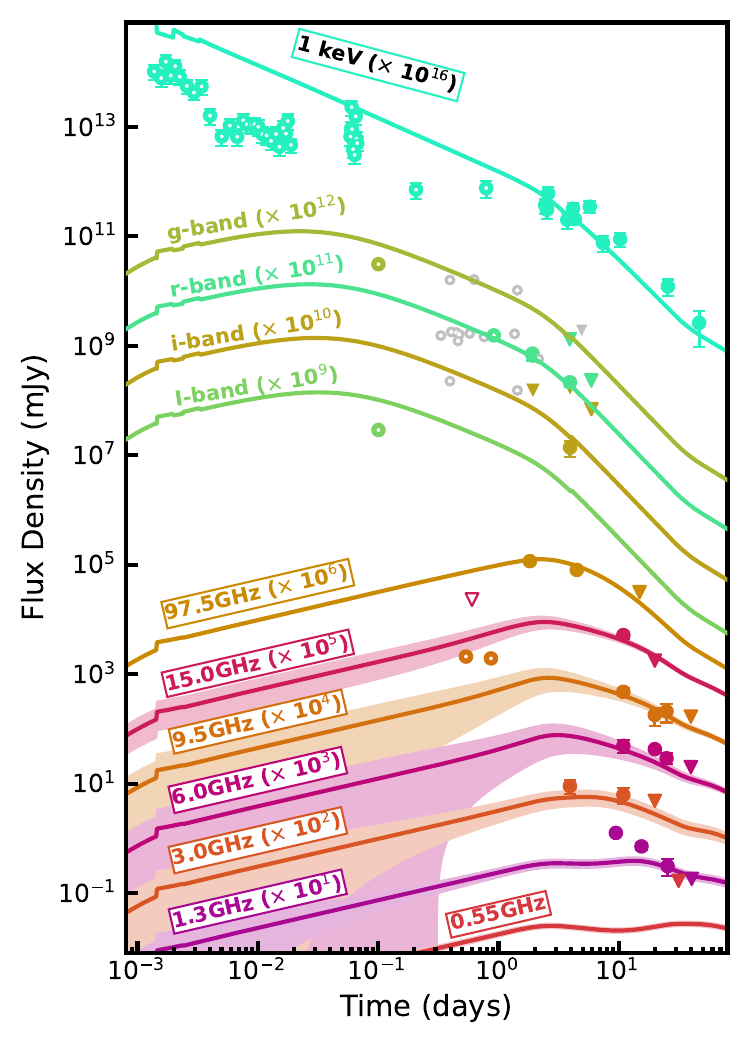}
    \caption{FS afterglow fit of \grb. Circles indicate detections while triangles show $3\sigma$ upper limits. Open symbols represent data masked in the afterglow fit ($\delta t \lesssim 1~$day). Gray points show comparison data from the literature (see Section~\ref{sec:Additional_Optical}), which we do not fit for in our model. Shaded regions represent the predicted variability of scintillation. It is clear that an FS model alone cannot account for the afterglow at $\lesssim 1~$day.}
    \label{fig:FSModel}
\end{figure}

We next fit the radio to X-ray afterglow of \grb\ within the modeling framework described by \citet{2014ApJ...781....1L}. This model samples the multidimensional parameter space of $p$, $E_{\rm K, iso}$, $n_0$, $\epsilon_{\rm e}$, $\epsilon_{\rm B}$, and $t_{\rm jet}$ using a Markov Chain Monte Carlo (MCMC) sampling with \texttt{emcee} \citep{2013ascl.soft03002F}. We require that $\epsilon_{\rm e} + \epsilon_{\rm B} < 1$, and assume that the fraction of participating  nonthermal (NT) electrons ($f_{\rm NT}$) is 1 \citep{2005ApJ...627..861E}. In addition to the standard FS framework, our modeling incorporates the scattering effects of scintillation \citep{1990ARA&A..28..561R, 2006ApJ...636..510G}, as well as the effects of inverse Compton (IC) cooling \citep{2001ApJ...548..787S, 2015ApJ...814....1L} and Klein Nishina (KN) corrections \citep{2009ApJ...703..675N, 2021MNRAS.504..528J, 2024arXiv240511309M}.

Since the data at $\delta t \lesssim 1~$days are not easily explained by a standard FS model (Section~\ref{sec:BasicConsiderations}), we exclude these from the model fits. 
We account for Milky Way dust extinction ($A_{V} = 0.196$~mag) and extinction along the line of sight ($A_{V, \rm GRB}$) assuming an SMC extinction law \citep{1992ApJ...395..130P}. In addition to the X-ray, optical, millimeter, and radio data presented in Section~\ref{sec:Observations_231117A}, we  incorporate the initial Arcminute Microkelvin Imager Large-Array (AMI-LA) nondetection reported by \citet{2023GCN.35142....1R} into our analysis. In the fit, we employ a 10\% uncertainty floor on our measurements.

We fit the X-ray to radio afterglow using 
128 walkers and 10,000 steps, and we discard the first $\sim 0.4\%$ of steps as ``burn-in'', over which the likelihood has not reached a stable value. Our best-fit model results in a $\chi^2/{\rm d.o.f.} \approx 82/82\approx 1.0$ and a highest log-likelihood value of $\mathcal{L} \approx 160$. We find $p \approx 2.29$, consistent with our expectations (Section~\ref{sec:BasicConsiderations}). The parameters of the best-fit model are $E_{\rm K, iso} \approx 2.7\times 10^{51}~$erg, $n_0 \approx 7.3 \times 10^{-2}~{\rm cm}^{-3}$, $\epsilon_{\rm e} \approx 0.97$, $\epsilon_{\rm B} \approx 7.2 \times 10^{-2}$, $t_{\rm jet} \approx 2.4~$days, and $A_{V, \rm GRB} \approx 0~$mag. We present our best-fit FS model in Figure~\ref{fig:FSModel} and the best-fit values and summary statistics in Table~\ref{tab:FS_fit}.

The ordering of the break frequencies is $\nu_{\rm sa} < \nu_{\rm m} < \nu_{\rm c}$ for the duration of our afterglow observations. As expected, $\nu_{\rm c} \approx \nu_{\rm X}$ for the duration of the detected X-ray afterglow.
At $\delta t \approx 8.4~$days, $\nu_{\rm mm} \approx \nu_{\rm m}$, as expected from the steepening of the mm light curve between $\delta t\approx 4.5$ and $14.7$~days. 
The optical afterglow is well fit within a factor of $\sim 3$ for the entirety of the afterglow, whereas the X-ray afterglow at $\delta t \lesssim 1~$days is overpredicted by a factor of $\sim 10$, as we expected. The $9.0$--$10.0~$GHz afterglow is underpredicted by a factor of $\sim 4$--$6$ at $\delta t \lesssim 1~$days; we return to this point in the next section. Furthermore, the $1.3~$GHz afterglow is underpredicted by a factor of $\sim 2$--$3$ at $\delta t \lesssim 15~$days, which is unsurprising given the observed flux density of $\sim 130~\mu$Jy at $\sim 9.5~$days is well in excess of our predicted $F_{\nu, \rm m}$ at a similar time ($\sim 50~\mu$Jy at $\sim 10.9~$days; Section~\ref{sec:BasicConsiderations}).

The measured $n_0$ for \grb is $\sim 14$ times higher than the median for short GRBs with $\epsilon_{\rm B} \approx 10^{-2}$ \citep[$\sim 5\times 10^{-3}~{\rm cm}^{-3}$;][ see also Section~\ref{sec:Discussion_231117A}]{2015ApJ...815..102F}.
We additionally find a best-fit value of $t_{\rm jet} \approx 2.4~$days, as expected, resulting in $\theta_{\rm j} \approx 10.4^\circ$, which is wider than the median for short GRBs with measured $\theta_{\rm j}$ \citep[$\sim 6.1^\circ$;][]{2023ApJ...959...13R}. The resulting beaming corrected kinetic energy is $E_{\rm K}\approx 4.4\times 10^{49}~$erg, consistent with the short GRB population \citep{2015ApJ...815..102F, 2023ApJ...959...13R}.

The derived value of $\epsilon_{\rm e}$ is higher than the expected equipartition value of $1/3$; however, high values have been found in several other afterglow studies \citep[e.g.,][]{2016ApJ...833...88L, 2018ApJ...858...65L, 2019MNRAS.489.2104T, 2021ApJ...911...14K, 2022ApJ...940...53S, 2022ApJ...935L..11L,
2023arXiv230810936S}. This tension can be alleviated by relaxing the assumption that $f_{\rm NT} = 1$. Indeed, a study of GRB afterglows found that $0.01 < \epsilon_{\rm e} < 0.2$ and $0.1 < f_{\rm NT}< 1$ \citep{2023MNRAS.518.1522D}, and
all of our fitted afterglow parameters are degenerate with $f_{\rm NT}$ (see, e.g., \citealt{2005ApJ...627..861E,
2022ApJ...935L..11L,  2023arXiv230810936S}). 

\begin{figure*}
    \centering
    \includegraphics[width = 0.9\textwidth]{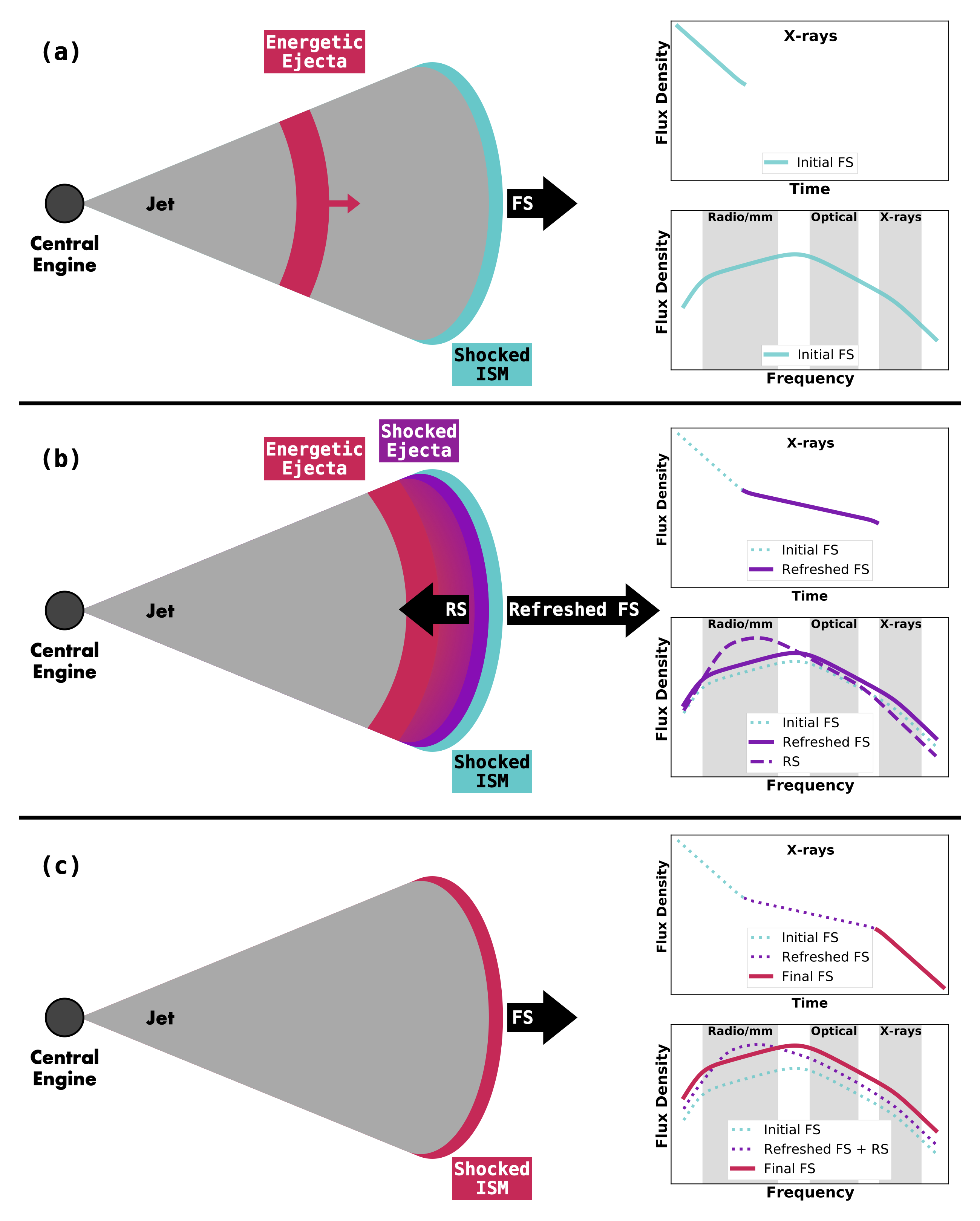}
    \caption{A schematic of the refreshed FS + RS model we use to describe the behavior observed in the afterglow of \grb. \emph{Top (a):} A central engine launches a jet of stratified ejecta. The fastest ejecta (blue) interact with the ISM and produce a FS (initial FS). Energetic ejecta (pink) are behind the fastest ejecta; no observational signature from these ejecta is detected yet. The observed light-curve behavior in the X-rays is a simple power-law decline. The observed spectrum is a triple broken power law, as expected for a synchrotron afterglow. \emph {Middle (b):} The energetic ejecta have caught up to the fastest ejecta, and deposit the energy into the FS (refreshed FS). The collision of these two ejecta shells shocks the ejecta (purple) and produces an RS that propagates backward. In the X-rays, the refreshed FS results in an observed plateau. In the afterglow spectrum, the RS dominates at lower frequencies, in the radio through optical. \emph {Bottom (c):} All of the energy has been deposited into the FS and the RS has dissipated. The X-rays continue a power-law decay, and the afterglow spectrum returns to a triple broken power law, both with increased energy (final FS).}
    \label{fig:Diagram}
\end{figure*}

\subsection{The Afterglow at $\lesssim 1~$days: A Refreshed Forward Shock with a Reverse Shock?}
\label{sec:EnjRS}

\begin{figure}
    \centering
    \includegraphics[width = 0.5\textwidth]{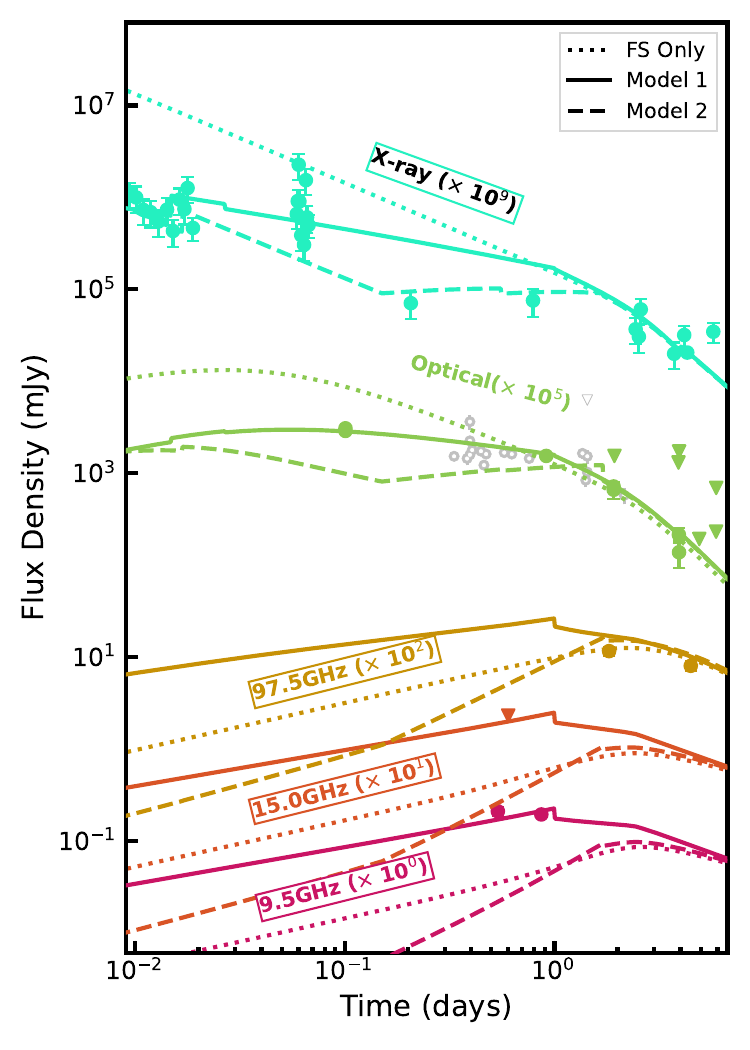}
    \caption{Comparison of the FS afterglow model (dotted lines) of \grb\ to the two suggested refreshed FS + RS models (solid lines and dashed lines), for selected wavelengths. Gray points represent data from the literature which are presented for comparison but not included in our models (Section~\ref{sec:Additional_Optical}).}
    \label{fig:RefreshedRSModel}
\end{figure}

We now explore the afterglow behavior at $\delta t \lesssim 1~$days. The flare in the X-ray afterglow at $\delta t \approx 0.06$~days may be attributed to late-time central engine activity, and the subsequent plateau in the X-ray afterglow is indicative of an energy injection episode \citep{2006ApJ...642..354Z, 2013MNRAS.430.1061R, 2015MNRAS.448..629G}. The leading physical explanations for energy injection are the transfer of braking radiation to the FS due to a millisecond magnetar central engine \citep{1998A&A...333L..87D, 2000ApJ...535L..33S, 2001ApJ...552L..35Z}, or a stratified jet with a distribution of ejecta Lorentz factors (see Figure~\ref{fig:Diagram}), leading to ejecta at lower Lorentz factors possessing a significant amount of energy catching up with the FS \citep{1998ApJ...496L...1R, 2000ApJ...535L..33S}. Regardless of the physical scenario causing the energy injection, the result is a refreshed FS from a gradual increase in kinetic energy, such that $E_{\rm K, iso} \propto t^{m}$, leading to an achromatic break in the afterglow light curve at the start ($t_0$) and end ($t_{\rm E}$) of the energy injection episode. We now explore an energy injection model to explain the observed afterglow behavior at $\delta t \lesssim 1~$day (excluding the X-ray flare, which is likely not dominated by the afterglow).

During the energy injection event, the expected flux evolution for $\nu_{\rm c} < \nu_{\rm obs}$ is $F_{\nu} \propto t^{(2-3p+m(2+p))/4}$ \citep{2006ApJ...642..354Z}. The X-ray light curve from $\delta t \approx 0.004$--$0.8~$days (ignoring the flaring event at $\delta t \approx 0.05$--$0.07~$days) can be fit with a single power law with a temporal index of $\alpha_{\rm X} = -0.6 \pm 0.1$, implying $m \approx 0.5$. This value of $m$ would lead to an optical decline of $\alpha = (3-3p+m(3+p))/4 \approx -0.2$ in the regime $\nu_{\rm m} < \nu_{\rm obs} < \nu_{\rm c}$ \citep{2006ApJ...642..354Z}, consistent with the observed plateau in the optical afterglow (Figure~\ref{fig:FSModel}).

While a refreshed FS can explain the X-ray and optical behavior at $\delta t \lesssim 1~$days, the original FS model already underpredicts the observed $9.0$--$10.0~$GHz afterglow at $\delta t \approx 0.5$--$0.9~$days, and the inclusion of energy injection will further reduce the predicted early radio flux, worsening the agreement.
This can be mitigated with the incorporation of a reverse shock (RS) to the model (see Figure~\ref{fig:Diagram}). An RS is a natural expectation from the interaction of the GRB jet with the surrounding ambient medium, or from an energy injection episode in the stratified jet model \citep{2000ApJ...535L..33S, 2000ApJ...532..286K, 2002ApJ...566..712Z, 2003ApJ...595..950Z, 2005ApJ...625..263G, 2005ApJ...628..315Z}. The broadband emission of several short GRBs has been explained with the invocation of refreshed FSs or RSs \citep{2006ApJ...650..261S, 2006MNRAS.372L..19F,2009ApJ...696.1871P, 2010MNRAS.409..531R, 2012A&A...541A..88H, 2013MNRAS.430.1061R, 2017ApJ...835...73Z, 2019ApJ...883...48L, 2019MNRAS.489.2104T, 2019ApJ...881...12B, 2021ApJ...906..127F, 2021ApJ...912...95R, 2023arXiv230810936S}; however, the combination of a refreshed FS with an RS has only been invoked to explain two other short GRBs \citep{2006ApJ...650..261S, 2019ApJ...883...48L}.  In the refreshed FS scenario, the RS will propagate during the injection period, and will naturally cross the ejecta at the cessation of energy injection \citep{1998ApJ...496L...1R, 2000ApJ...532..286K}, potentially producing detectable synchrotron emission \citep{2000ApJ...535L..33S, 2011ApJ...733...86U}.

The RS can be described by its own set of break frequencies ($\nu_{\rm sa, RS}, \nu_{\rm m, RS}, \nu_{\rm c, RS}$) and peak flux ($F_{\nu, \rm m, RS}$). At $t_{\rm E}$, the RS break frequencies are expected to be related to the FS break frequencies ($\nu_{\rm sa, FS}, \nu_{\rm m, FS}, \nu_{\rm c, FS}$) and peak flux ($F_{\nu, \rm m, FS}$) such that
\begin{align} 
    \frac{\nu_{\rm m, RS}}{\nu_{\rm m, FS}} &\sim \Gamma^{-2} R_{\rm B}\, , \nonumber \\
    \frac{\nu_{\rm c, RS}}{\nu_{\rm c, FS}} &\sim  R_{\rm B}^{-3}\, ,\nonumber \\
    \frac{F_{\nu, \rm m, RS}}{F_{\nu, \rm m, FS}} &\sim \Gamma R_{\rm B}\, ,
\end{align}

\noindent where $R_{\rm B} = (\epsilon_{\rm B, RS}/\epsilon_{\rm B, FS})^{1/2}$ is the ejecta magnetization and $\Gamma$ is the bulk Lorentz factor of the FS at $t_{\rm E}$ \citep{2003ApJ...595..950Z, 2018ApJ...862...94L}. We assume $\nu_{\rm sa, RS}\propto \Gamma^{8/5}{\nu_{\rm sa, FS}}$ \citep{2000ApJ...535L..33S} and that the RS and FS have similar values of $p$, $\epsilon_{\rm e}$, and Compton $Y$ parameters. For a Newtonian RS, the evolution of the RS break frequencies after $t_{\rm E}$ is dependent on the ejecta Lorentz factor evolution, such that $\Gamma \propto R^{-g}$ (where $R$ is the radius of the ejecta shell; \citealt{1999MNRAS.306L..39M, 2000ApJ...542..819K}). For an ISM environment, numerical simulations indicate $g \approx 2.2$ \citep{2000ApJ...542..819K}.

To test this model, we construct a refreshed FS $+$ RS light-curve model \citep[see][]{2018ApJ...862...94L}. We set $m = 0.5$, $t_0 = 0~$days\footnote{This model effectively starts at panel (b) in Figure~\ref{fig:Diagram}.}, and $t_{\rm E} = 1.0~$days, at which time $\Gamma \approx 4.7$. In order to ensure consistency with the FS parameters, we scale the RS break frequencies and peak flux according to the prescription laid out above, and set $g = 2.2$. We find that $R_{\rm B} = 0.3$ provides a good fit\footnote{The resulting RS parameters at $\delta t = 1.0~$days are $\nu_{\rm sa, RS} \approx 6 \times 10^8~{\rm Hz}$, $\nu_{\rm m, RS} \approx 6\times 10^{10}~{\rm Hz}$, $\nu_{\rm c, RS} \approx 7\times 10^{18}~{\rm Hz}$, and $F_{\nu, \rm m, RS} \approx 0.4~{\rm mJy}$.} to the $9$--$97.5~$GHz data at $\delta t \lesssim 2$~days (``Model 1'', Figure~\ref{fig:RefreshedRSModel}). While this model fits the majority of the data, the refreshed FS overpredicts the X-ray plateau at $\delta t \approx 0.2$--$0.8~$days by a factor of $\sim 3$--$5$. Additionally, given $\nu_{\rm m, RS} \ll \nu_{\rm O}$, the RS does not contribute significantly to the optical afterglow, and therefore we still are unable to explain our measured $\beta_{\rm O}$ at $\delta t \approx 1.1~$days.

We therefore explore an alternative refreshed FS $+$ RS model to better fit the X-ray plateau. To match $\alpha_{\rm X} \approx 0$ at $\delta t \approx 0.2$--$0.8~$days, we require a higher value of $m \approx 0.9$, as well as $t_{0} > 0~$days to fit the early ($\delta t \approx 4\times 10^{-3}$--$2\times 10^{-2}~$days) X-rays.
Physically, this may indicate that the RS is a result of a violent collision of a secondary ejecta shell, released after the initial GRB ejecta \citep[e.g.,][]{2018ApJ...859..134L, 2023arXiv230810936S}. The X-ray flare at $\delta t \approx 0.06~$days may indicate the release of this shell, which must then catch up with the initial shell. 
However, since the optical afterglow at $\delta t \lesssim 1~$days will be underpredicted by a factor of $\sim 2$--$9$, we compensate for this by adjusting the RS break frequencies and peak flux\footnote{As $t_0 \neq 0$, it is not required that the RS from a shell collision must be consistent with the FS.}. We find that a refreshed FS $+$ RS model with $f_{\nu, \rm m, RS} \approx 0.2~$mJy, $\nu_{\rm m, RS} \approx 10^{13}~$Hz, $\nu_{\rm c, RS} \approx 4 \times 10^{13}~$Hz, $m = 0.9$, $t_0 = 0.15~$days, and $t_{\rm E} = 1.7~$days provides the most adequate match to the X-ray plateau, within a factor of $\sim 1.3$ (``Model 2'', Figure~\ref{fig:RefreshedRSModel}). We note that $\nu_{\rm sa, RS}$ is unconstrained in this model, and we set it to  $\sim 10^{9}~$Hz, similar to Model 1. 

In Model 2, $\nu_{\rm m, RS} \approx \nu_{\rm O}$ at $\delta t \lesssim 1.7~$days, and therefore the optical afterglow at $\delta t \approx 1.1~$days is significantly affected by the RS. As a result, $\beta_{\rm O}$ at $\delta t \lesssim t_{\rm E}$ is steeper than the FS expectation of $\sim -0.65$, and we find $\beta_{O} \approx -0.74$ at $\sim 1.1~$days for Model 2, though this value is still not as steep as our measured $\beta_{\rm O} = -0.86 \pm 0.01$ (Section~\ref{sec:ag-subtraction}). Additionally, the combination of the later $t_0$ and higher $m$ (compared to Model 1), results in an underestimate of the optical afterglow at $\delta t \approx 0.1~$days by a factor of $\sim 3$--$5$. Owing to our given $\nu_{\rm m, RS}$, the RS does not produce significant flux at $9.0$--$10.0~$GHz, and therefore the radio afterglow at $\delta t \lesssim 1~$days remains underpredicted in Model 2.

Both refreshed FS $+$ RS models can each explain individual components of the broadband emission, but it is ultimately challenging to explain the X-ray plateau self-consistently with the other bands. Model 1 moderately overpredicts the early X-ray plateau, whereas Model 2 better predicts the plateau within a factor of $\sim 1.3$. Conversely, Model 2 is unable to explain the $9.0$--$10.0~$GHz afterglow at $\lesssim 1~$days, whereas the RS in Model 1 naturally explains this emission. Additionally, while Model 2 may alleviate some of the tension between our measured and expected $\beta_{\rm O}$ at $\delta t \approx 1.1~$days, it more significantly underpredicts the optical afterglow at $\delta t \approx 0.1~$days, whereas Model 1 matches the optical afterglow at $\delta t \approx 0.1~$days within a factor of $\sim 2$. Given that Model 1 is consistent with the FS in terms of RS parameters, we moderately prefer this scenario over Model 2, which requires more ``fine-tuned'' RS parameters. In both refreshed FS $+$ RS scenarios presented above, $E_{\rm K, iso}$ increases by a factor of $\sim 10$ over the time span of $\delta t \approx 10^{-2}$--$1$~days, indicating that a significant amount of energy was released at later times and/or in slower moving ejecta. If the energy injection is the result of a millisecond magnetar central engine, both models would require central engine activity out to $\gtrsim 1~$day, which is much longer than the expected spin-down timescale \citep[e.g.,][]{2011MNRAS.413.2031M, 2014MNRAS.439.3916M}. Therefore, we expect the energy injection to be the result of stratified ejecta, which would not require central engine activity at very late times.

\section{The Properties of Radio-detected Short GRBs}
\label{sec:Discussion_231117A}

\grb is the second short GRB to be detected with MeerKAT \citep{2023arXiv230810936S} and ALMA \citep{2022ApJ...935L..11L}, and the first short GRB to be detected by both facilities. Our radio/mm afterglow detections span nearly two decades in frequency ($\nu = 1.3$--$97.5~$GHz), essential for constraining both the properties of the refreshed FS and the RS (Section~\ref{sec:AGModeling}). These detections motivate us to explore the properties of radio-detected short GRBs to understand which properties contribute to our ability to detect these events at radio bands ($\sim 1$--$20~$GHz). 

Broadening our discussion to all short or merger-driven GRBs, we explore whether the radio-detected population differs from the overall short GRB population in terms of their redshifts, X-ray afterglows, circumburst densities, projected offsets from their host galaxies, and opening angles. We also include GRBs that have been classified, or tentatively classified, as short GRBs with extended emission \citep{2016ApJ...829....7L, 2022ApJ...940...56F} and exclude short GRBs with likely single-star progenitors \citep[e.g., GRBs\,080913, 090423, 100724A, and 200826A;][]{2009ApJ...703.1696Z,  2010GCN.10976....1U, 2021NatAs...5..911Z, 2021NatAs...5..917A, 2021MNRAS.503.2966R}. 
We finally include three merger-driven long GRBs (GRB\,060614A, GRB\,211211A, and GRB\,230307A; \citealt{2006Natur.444.1050D, 2006Natur.444.1047F, 2006Natur.444.1053G,  
2006Natur.444.1044G, 2015NatCo...6.7323Y, 2022Natur.612..223R, 2022Natur.612..228T, 2022Natur.612..232Y, 2024Natur.626..737L, 2024Natur.626..742Y}). We further require an X-ray afterglow detection, resulting in a sample of 132 short/merger driven GRBs.
Of this total sample, 89 have redshifts, 57 have measured circumburst densities, 84 have measured projected offsets from their host galaxies, and 32 have measured $\theta_{\rm j}$ or published lower limits ($\theta_{\rm j, lim}$).

Within our sample of 132 GRBs, 71 were observed at radio wavelengths ($\sim 1$--$20~$GHz) and 17 were detected in the radio band, including the merger-driven GRB\,230307A \citep[][this work]{2005Natur.438..988B, 2006ApJ...642..354Z, 2014ApJ...780..118F, 2015ApJ...815..102F, 2019ApJ...883...48L, 2019MNRAS.489.2104T, 2021ApJ...906..127F, 2022ApJ...935L..11L, 2024Natur.626..737L}. 
For the purposes of this discussion, we label short GRBs with radio nondetections $\lesssim 100~\mu$Jy as those with ``deep'' radio limits, and those with nondetections $\gtrsim 100~\mu$Jy as ``shallow'' radio limits.

With 17 (16 short +1 merger-driven) GRBs with radio detections, $\sim 13\%$ of our sample has been detected at radio wavelengths. Comparatively,  only $\sim 7\%$ of the \citet{2015ApJ...815..102F} X-ray-detected sample were detected in the radio. With this larger and more statistically significant sample size, we finally have the capability to look at the properties of radio-detected short GRBs, compared to their undetected counterparts.

\begin{figure*}
    \centering
    \includegraphics[width = \textwidth]{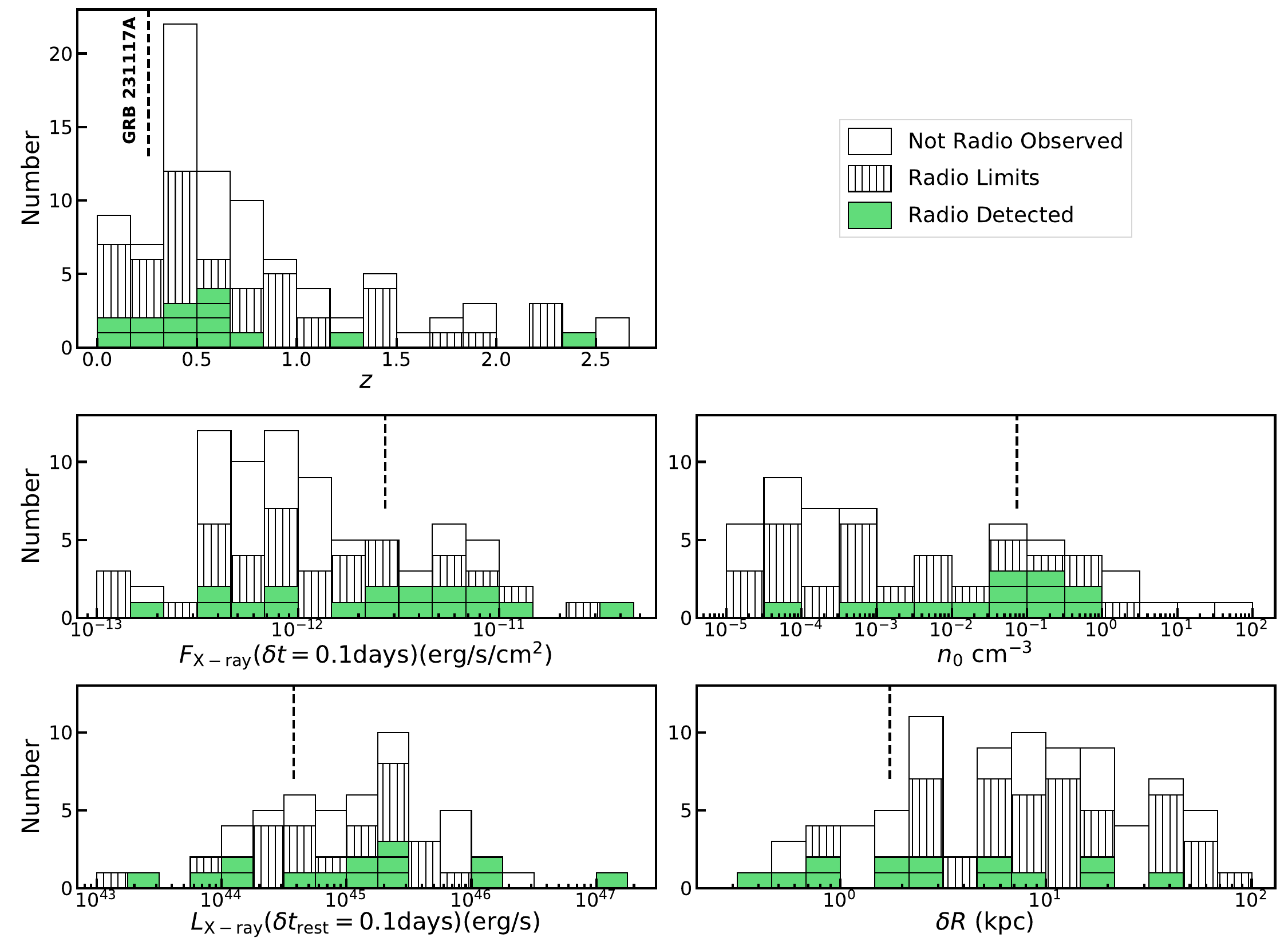}
    \caption{Distributions of the short GRB population for the following properties: redshift ($z$, top left), X-ray flux at $\delta t = 0.1~$days ($F_{\rm X, \delta t = 0.1 days}$, middle left), X-ray luminosity at $\delta t_{\rm rest} = 0.1~$days ($L_{\rm X, \delta t_{\rm rest} = 0.1 days}$, bottom left), circumburst density ($n_0$, middle right), and projected offset from host galaxy ($\delta R$ in kpc, bottom right). The colored distribution represents short GRBs with radio detections. For each property, we indicate the location of \grb with a dashed line. The white, unhatched distribution represents short GRBs that were not observed in the radio. The hatched distribution represent GRBs with radio limits.}
    \label{fig:Histograms}
\end{figure*}

\subsection{Redshift}

We first explore the redshift distribution of the short GRB population. The brightness of the afterglow of a GRB follows the inverse-square law of distance;  therefore, it is expected that higher redshift GRBs will have fainter radio afterglows, potentially leading to a tendency for radio-detected GRBs to be observed at lower redshifts \citep[e.g.,][]{2017ApJ...835...73Z}.
In total, there are 87 short GRBs with redshift determinations\footnote{For 9 short GRBs, we use the photometrically derived redshifts from \citet{2024ApJ...962....5N}, using the mass-radius relation.} \citep{2009GCN..9958....1L, 2022MNRAS.515.4890O, 2022ApJ...940...56F, 2023arXiv230810936S, 2024Natur.626..737L, 2024ApJ...962....5N}, and 14 of those short GRBs have radio afterglow detections\footnote{GRB\,201006A, GRB\,230205A, and GRB\,230217A do not have redshift determinations.} (Figure~\ref{fig:Histograms}, top left). At $z = 0.257$, \grb is among the most nearby short GRBs, with only $12/87$ short GRBs at a lower redshift, whereas the total short GRB population has a median redshift of $\langle z_{\rm total} \rangle = 0.6$. In general, the short GRBs that are detected in the radio reside at slightly lower redshifts, with a median value of $\langle z_{\rm radio} \rangle = 0.5$. However, the observational bias is more apparent from a comparison at higher redshifts: only 3 ($\sim 18\%$) of the radio-detected short GRBs have been found at $z > 0.6$, compared to half of the total short GRB population. This apparent bias toward low-redshift short GRBs confirms our expectation that it is more difficult to detect the radio afterglow at high redshifts.
Importantly, deep radio limits of short GRBs exist across the redshift range, indicating that we are not biasing our radio observations toward low-redshift short GRBs.

\subsection{The X-ray Afterglow}

We next explore trends with the X-ray afterglow fluxes and luminosities. This comparison is rooted in long GRBs, as a study of their radio properties \citep{2012ApJ...746..156C} found that their radio detectability is correlated to the X-ray afterglow flux;  we therefore would expect a similar pattern for short GRBs.
Thanks to the automatic slewing of {\it Swift}/XRT, the majority ($\sim 70\%$) of short GRBs have been detected in the X-rays \citep{2015ApJ...815..102F, 2023ApJ...959...13R}, resulting in a large afterglow sample. We first download the X-ray light curves from the {\it Swift} website, and supplement the X-ray light curves with additional published data \citep{2005Natur.437..845F, 2023ApJ...959...13R, 2024Natur.626..737L, 2024Natur.626..742Y}.

In order to compare the X-ray afterglows of short GRBs, we determine the X-ray flux at a common time of $\delta t = 0.1~$days ($F_{\rm X, 0.1 \ days}$). We choose this time because typically the X-ray afterglow has settled into its power-law decay on this timescale \citep{2006ApJ...642..354Z, 2007RSPTA.365.1213B, 2010MNRAS.406.2149M} but is still observable and detected. Of the 132 short GRBs in our sample, we discard 53 owing to insufficient X-ray data: the X-ray afterglow is not detected beyond $\delta t = 0.1~$days, and/or there is only one X-ray detection. Of the remaining GRBs, 5 have X-ray observations starting at $> 0.1~$days, and we therefore determine $F_{\rm X, 0.1 days}$ via extrapolation, assuming the afterglow behaved like a single power law. 
For the remaining bursts, we use the closest X-ray data point at $\delta t =0.1 \pm 0.03~$days for $F_{\rm X, 0.1 days}$. If there is not a data point within that time range, we instead interpolate the X-ray light curve to $\delta t = 0.1~$days using the X-ray detections surrounding our time of interest. 
In total, we find $F_{\rm X, 0.1 \ days}$ measurements for 79 short GRBs, including all 17 short GRBs with radio detections (Figure~\ref{fig:Histograms}, middle left). Similarly, for the 52 events with redshifts and sufficiently sampled X-ray afterglows, we determine their X-ray luminosities at a common rest-frame time of $\delta t_{\rm rest} = 0.1~$days ($L_{\rm X, 0.1 days}$). This includes 14 short GRBs with radio detections (Figure~\ref{fig:Histograms}, bottom left).

The radio-detected population has a median X-ray flux of $\langle F_{\rm X, 0.1 days}\rangle \approx 2.7 \times 10^{-12}~{\rm erg~s}^{-1}{\rm cm}^{-2}$, $\sim 3$ times higher than the overall short GRB population ($\langle F_{\rm X, 0.1~days}\rangle \approx 8.1 \times 10^{-13}~{\rm erg~s}^{-1}{\rm cm}^{-2}$).
The higher median for radio-detected afterglows is expected, as both the radio and X-ray afterglows are scaled by $F_{\nu, \rm m}$, and thus a brighter X-ray afterglow implies a higher $F_{\nu, \rm m}$ and therefore a brighter radio afterglow.

Despite the fact that the detection of a radio afterglow selects for lower redshift and high $F_{\rm X, 0.1 \ days}$, we find that radio-detected short GRBs have similar X-ray luminosities to the overall population ($\langle L_{\rm X, 0.1 \ days}\rangle \approx 1.5 \times 10^{45}~{\rm erg/s}$ for the radio-detected events compared to the population median of $\langle L_{\rm X, 0.1 \ days}\rangle \approx 1.1 \times 10^{45}~{\rm erg/s}$).  The X-ray afterglow of \grb\ is in the lower 1/3 of the short GRB population in terms of X-ray luminosity, with $L_{\rm X, 0.1 \ days} \approx 3.8 \times 10^{44}$~erg~s$^{-1}$. Overall, radio-detected short GRBs span $\sim 4$ orders of magnitude in X-ray afterglow luminosity, indicating that there is no clear correlation between presence of radio afterglow and intrinsic afterglow luminosity, consistent with the findings of \citet{2017ApJ...835...73Z}.

\subsection{Local Environment and Location Within the Host}

We next explore the distribution of circumburst densities for short GRBs. At radio wavelengths, the afterglow brightness is strongly dependent on $n_0$ ($F_\nu \propto n_0^{1/2}$ for $\nu_{\rm sa} < \nu_{\rm obs} < \nu_{\rm m}$; \citealt{2002ApJ...568..820G}), and we therefore expect that the radio-detected population should have higher circumburst densities. In addition to that of \grb, we gather $n_0$ measurements for 52 short GRBs and 3 merger-driven long GRBs \citep{2009ApJ...696..971X, 2015ApJ...815..102F, 2022Natur.612..223R, 2023ApJ...959...13R, 2024Natur.626..737L}. In the cases where multiple $n_0$ solutions are derived for a GRB \citep[e.g.,][]{2015ApJ...815..102F}, we choose the $n_0$ corresponding to the lower value of $\epsilon_{\rm B}$\footnote{While this results in higher measured $n_0$ values, we choose these values as $\epsilon_{\rm B} < 0.1$ is supported by broadband afterglow models of short and merger driven GRBs \citep[e.g.,][]{2019MNRAS.489.2104T, 2021ApJ...906..127F, 2022Natur.612..223R, 2022ApJ...935L..11L, 2023arXiv230810936S, 2024Natur.626..737L}}. For two bursts, GRB\,210726A and GRB\,211106A, we use values based on updated redshifts \citep{2022ApJ...935L..11L, 2023arXiv230810936S, 2023A&A...678A.142F, 2024ApJ...962....5N}.
Overall, our sample contains 57 GRBs\footnote{This includes 12 GRBs with an unknown redshift, in which case the redshift was either assumed to be 0.5 \citep{2015ApJ...815..102F} or 0.64 \citep{2023ApJ...959...13R}.} with measured $n_0$, including 13 short GRBs with radio detections (Figure~\ref{fig:Histograms}, middle right).

As expected, the difference between radio-detected short GRBs and the total short GRB population is starkly apparent in $n_0$. The median $n_0$ for radio-detected short GRBs is over a magnitude higher than for the total short GRB population\footnote{We find this trend holds even when accounting for the changes in the measured $n_0$ due to assumptions of $\epsilon_{\rm B}$.}
($\langle n_{0, \rm radio}\rangle \approx 7.4 \times 10^{-2}~{\rm cm}^{-3}$ compared to $\langle n_{0, \rm total}\rangle \approx 1.9 \times 10^{-3}~{\rm cm}^{-3}$).
The circumburst density of \grb is at the median for radio-detected short GRBs, at $n_0 \approx 7.3\times 10^{-2}~{\rm cm}^{-3}$. Furthermore, we find that $9/16$ ($\sim 56\%$) radio-observed short GRBs
with high $n_0 \gtrsim 10^{-2}~{\rm cm}^{-3}$ resulted in detection, compared to only $4/23$ ($\sim 17\%$) of the radio-observed short GRBs with low $n_0 \lesssim 10^{-2}~{\rm cm}^{-3}$.
Overall, the radio-detected population of short GRBs is clearly biased toward higher circumburst densities.

As we find radio-detected short GRBs in higher $n_0$ environments, we would naively expect that these GRBs trace smaller projected offsets from their host galaxies. Thus, we explore the distribution of projected offsets for short GRBs.
We gather the projected offsets in kpc for 84 short/merger-driven GRBs with known redshifts \citep[][this work]{2022ApJ...940...56F, 2024ApJ...962....5N, 2023arXiv230810936S, 2022Natur.612..223R, 2024Natur.626..737L}, 14 of which are radio-detected afterglows (Figure~\ref{fig:Histograms}, bottom right).
As expected, we find the radio-detected population, including \grb, resides at lower projected offset ($\langle \delta R_{\rm radio} \rangle \approx 2.7~$kpc) compared to the total GRB population ($\langle \delta R_{\rm total} \rangle \approx 7.8~$kpc). Indeed, of the 8 ($\sim 10\%$) short GRBs that reside at $< 1~$kpc, 4 have radio detections and 3 of the remaining 4 GRBs were not even observed at radio wavelengths. At the high-offset end,
only 4 of the 14 radio-detected short GRBs ($\sim 24\%$) reside at $\delta R \gtrsim 8~$kpc, compared to half of the short GRB population overall. This indicates that radio-detected short GRBs are more centrally concentrated, commensurate with their higher densities.

\begin{figure}
    \centering
    \includegraphics[width = \columnwidth]{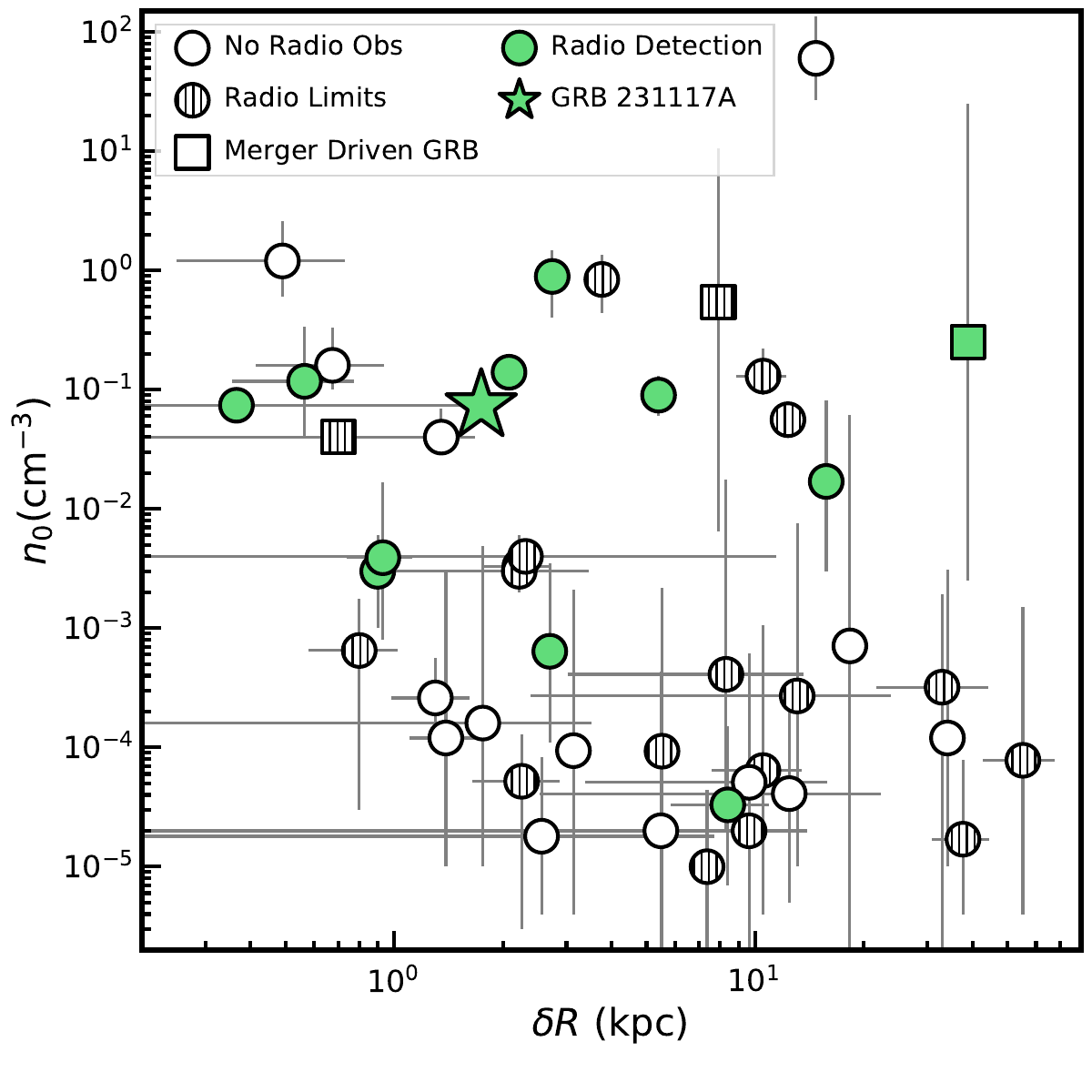}
    \caption{Circumburst density ($n_0$) vs. projected offset from the host galaxy ($\delta R$) for short GRBs. Filled-in points represent short GRBs with radio detections. \grb is represented as a star. Empty points represent short GRBs that were not observed at  radio wavelengths. Hatched points represent GRBs with radio limits.}
    \label{fig:Correlation}
\end{figure}

If short GRBs trace the typical ISM density of their host galaxies, we expect that $n_0$ would decrease as a function of offset. Early studies based on smaller numbers of measurements (22 events;  \citealt{2015ApJ...815..102F}) found no obvious correlation between the two parameters. Now with a sample of 45 GRBs,  we indeed find a trend that $\delta R$ and $n_0$ are inversely related, with GRBs at smaller projected offsets having higher measured $n_0$ values (Figure~\ref{fig:Correlation}). Specifically, there is a clear dearth of short GRBs at low $n_0 \lesssim 10^{-2}~{\rm cm}^{-3}$ at a projected offset of $\lesssim 1~$kpc. This dearth can be seen in Figure~9 of \citet{2015ApJ...815..102F}; however, given our larger sample size the trend is more apparent. Of the $5$ short GRBs at $\delta R \lesssim 1~$kpc with high measured $n_0$ ($> 10^{-2}~{\rm cm}^{-3}$), $2$ are radio-detected, whereas two were not observed at radio wavelengths, and one (GRB\,060614A) was observed but not detected (Figure~\ref{fig:Correlation}). Based on the observed trend, the two unobserved events would be strong candidates for radio detection, and we return to the single undetected GRB in Section~\ref{sec:Whynotdetected}.

\subsection{Opening Angles}

\begin{figure}
    \centering
    \includegraphics[width = \columnwidth]{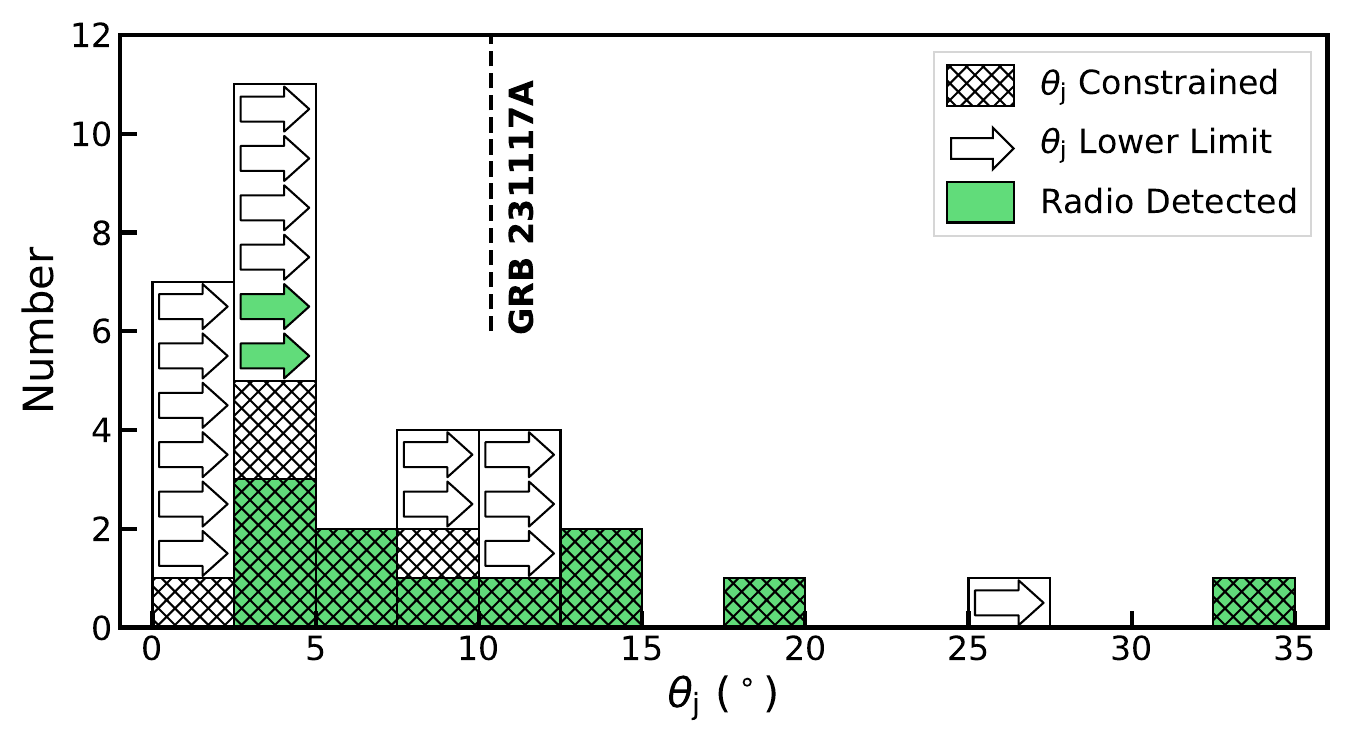}
    \caption{Distribution of the jet opening angle ($\theta_{\rm j}$) or lower limit on the opening angle ($\theta_{\rm j, lim}$). The green distribution represents radio detection short GRBs. The hatched distribution represents short GRBs with a constrained $\theta_{\rm j}$, and the empty distribution with arrows represents short GRBs with a measured $\theta_{\rm j, lim}$.}
    \label{fig:theta_j}
\end{figure}

We next explore the distribution of $\theta_j$ for short GRBs. Understanding the true $\theta_{\rm j}$ distribution is important for determining the true short GRB rate \citep{2015ApJ...815..102F, 2018ApJ...857..128J, 2020MNRAS.492.5011D, 2023ApJ...959...13R}, as well as comparing the rate to the neutron star merger rate observed by gravitational-wave observatories such as LIGO-Virgo-KAGRA \citep{2020ApJ...892L...3A, 2023PhRvX..13d1039A} to determine what fraction of mergers launch short GRBs. Recent radio detections of short GRBs have revealed wide ($> 10^\circ$) jets \citep{2022ApJ...935L..11L, 2023arXiv230810936S}, suggesting that without radio observations, we may be biasing our $\theta_{\rm j}$ distribution toward narrow ($< 10^\circ$) jets.
We use the 29 uniformly derived $\theta_j$ values presented by \citet{2023ApJ...959...13R}, which encompasses the largest such study for short GRBs. 
When relevant, we use updated $\theta_j$ measurements for GRBs\,060614A, 210726A, 211106A, 211211A, and 230307A \citep{2009ApJ...696..971X, 2023arXiv230810936S, 2022ApJ...935L..11L, 2022Natur.612..223R, 2024Natur.626..737L}, as well as GRB\,231117A (this work). Our sample includes 15 measurements (18 limits), 11 (2) of which have radio-afterglow detections (Figure~\ref{fig:theta_j}). While this sample comprises the fewest bursts out of our subsamples, it has the highest fraction of radio-detected short GRBs ($13/32 \approx 40\%$).  This not only demonstrates the difficulty of placing meaningful constraints on $\theta_{\rm j}$, but also indicates that radio detections are important for measuring $\theta_{\rm j}$.

The radio-detected population spans the majority of the $\theta_{\rm j}$ distribution ($\theta_{\rm j} \approx 3$--$34^{\circ}$), indicating there is no obvious bias for a short GRB to be detected at radio wavelengths based on the collimation of the jet. However, all of the short GRBs with measured $\theta_{\rm j} > 10^\circ$, including \grb, are detected in the radio, indicating that radio detections are important for constraining $\theta_{\rm j}$ for wider jets. In order to measure $\theta_{\rm j}$, a jet break must be detected in one or more bands. As wider jets ($> 10^\circ$) correspond to later $t_{\rm jet}$ ($\theta_{\rm j} \propto t_{\rm jet}^{3/8}$; \citealt{1999ApJ...519L..17S}), the X-ray and optical afterglow may fade beyond our detection limits prior to $t_{\rm jet}$. Therefore, radio observations are sometimes the only reliable way to determine $t_{\rm jet}$, as radio afterglow is expected to rise until $\delta t = t_{\rm jet}$ (unless $\nu_{\rm m} < \nu_{\rm R}$). Indeed, in the cases of the wide jets of GRB\,210726A and GRB\,211106A, the detection of a jet break was only uncovered by the radio afterglow \citep{2022ApJ...935L..11L, 2023arXiv230810936S}, and X-ray analysis alone would have only led to a lower limit on $\theta_{\rm j}$ \citep{2023ApJ...959...13R}.

\subsection{GRBs that Evade Radio Detection}
\label{sec:Whynotdetected}

We have demonstrated that the radio-detected population tends to be found at low redshift ($z < 0.6$), high $F_{\rm X, 0.1 \ days}$ ($\gtrsim 10^{-12}~{\rm egs~s}^{-1}{\rm cm}^{-2}$), high density ($n_0 \gtrsim 10^{-2}~{\rm cm}^{-3}$), and low projected offset ($\delta R \lesssim 8~$kpc). These properties are consistent with expectations from theoretical predictions of afterglows \citep[e.g.,][]{2002ApJ...568..820G}, as well studies of radio-detected long GRBs \citep[e.g.,][]{2012ApJ...746..156C}, indicating that the standard FS model describes the afterglow behavior of most short GRBs. Here, we investigate the short GRBs with radio limits to understand if there are events that go against expectations set by the rest of the population. 

Among all of the subsamples defined by their measured properties, we find that $n_0$ is the strongest indicator as to whether a short GRB will have a detectable radio afterglow. There exist 5 short GRBs and 2 merger-driven GRBs with high $n_0 > 10^{-2}~{\rm cm}^{-3}$ that were observed but not detected in the radio band: GRBs\,050709A, 060614A, 070714B, 090621B, 110112A, 111117A, and 211211A. For GRBs\,090621B, 110112A, and 111117A, each burst was only observed at $\sim 5$ or $\sim 8~$GHz in one epoch at $\delta t < 2~$days \citep{2012ApJ...756...63M, 2015ApJ...815..102F}. Similarly, GRB\,070714B was only observed at $\sim 8~$GHz in one epoch at $\delta t \approx 15.6~$days \citep{2015ApJ...815..102F}. As demonstrated in Figure~\ref{fig:RadioAGCompilation}, many radio afterglows do not peak until $\delta t > 2~$days, and fade beyond detection by $\delta t < 15~$days. Therefore, the lack of radio detections for these 4 GRBs may be attributed to insufficient temporal spacing of radio observations. Additionally, all 4 of these GRBs have $F_{\rm X, 0.1~days} \lesssim 10^{-12}~{\rm egs~s}^{-1}{\rm cm}^{-2}$, and 2 (GRBs\,070714B and 111117A) are at $z > 0.6$ \citep{2022ApJ...940...56F}. We have demonstrated that we are less efficient at detecting bursts with these properties.

The remaining 3 GRBs are GRB\,050709A, 060614A, and 211211A. 
The lack of radio detection for both GRB\,060614A and GRB\,211211A is easily explained by a narrow jet ($\theta_{\rm j} \lesssim 5^{\circ}$), resulting in an early $t_{\rm jet} \lesssim 1.3~$days \citep{2007A&A...470..105M, 2009ApJ...696..971X, 2022Natur.612..223R}. Both of these bursts were only observed in the radio at $\gtrsim 6~$days \citep{2006GCN..5359....1L, 2022Natur.612..223R}, and therefore 
it is likely that we missed any detectable radio emission owing to the timing of the observations.
However, the radio observations of GRB\,050709A are well spaced (four epochs spanning $\delta t \approx 1.6$--$7.5~$days) and sufficiently deep (reaching $F_{\nu} \lesssim 40~\mu$Jy). Accordingly, the lack of radio detections for GRB\,050709A is difficult to explain for many afterglow models \citep[e.g.,][]{2005Natur.437..845F, 2006MNRAS.367L..42P}. Additionally, both \citet{2015ApJ...815..102F} and \cite{2023ApJ...959...13R} derive high measured $n_0 \approx 1~{\rm cm}^{-3}$, though afterglow models with lower values of $n_{0} \lesssim (1$--5) $ \times 10^{-2}~{\rm cm}^{-3}$ are still feasible \citep{2005Natur.437..845F, 2006MNRAS.367L..42P, 2024arXiv240402627O}. Overall, GRB\,050709A is the only short GRB that does not seem to fit the standard FS model assumptions given the lack of radio detection.

\subsection{Prospects for Future Radio Observations}

We now explore if radio detectability of short GRB afterglows can be improved by utilizing the learned trends between radio afterglows and other observables. For instance, there is a relatively high radio detection rate for short GRBs with higher X-ray fluxes of $F_{\rm X,0.1~days} > 10^{-12}~{\rm egs~s}^{-1}{\rm cm}^{-2}$, detecting $11$/23 radio-observed short GRBs ($\sim 48\%$), compared to only 6/27 ($\sim 27\%$) short GRBs with lower fluxes of $F_{\rm X,0.1~days} < 10^{-12}~{\rm egs~s}^{-2}{\rm cm}^{-2}$. Furthermore, of the 71 short GRBs with radio observations, the first observation commenced at $\delta t > 0.5~$days for $\sim 74\%$ of the events, at which point $F_{\rm X,0.1~days}$ was known. We suggest that one way to  maximize the chance of radio detection is to increase the depth of our observations for short GRBs with $F_{\rm X, 0.1~days} < 10^{-12}~{\rm egs~s}^{-1}{\rm cm}^{-2}$.

Additionally, our findings may lend insight into future host associations and observations. We have found that radio-detected short GRBs are often found in high-density environments at low offsets from their host galaxies, including \grb.
An interesting case study is GRB\,210726A, which was initially associated with an galaxy at $z = 0.35 \pm 0.15$ with a angular offset of $\sim 3.1''$ \citep{2021GCN.30534....1W}, resulting in a physical offset of $\sim 15^{+4}_{-5}~$kpc. However, deeper observations revealed a faint underlying host galaxy at a photometric redshift of $z \approx 2.4$ \citep{2023arXiv230810936S}, resulting in a much lower angular offset of $\sim 0.04''$, or a physical offset of $\sim 0.4~$kpc \citep{2022ApJ...940...56F}. Similar galaxy associations resulted from radio detections of GRB\,211106A \citep{2022ApJ...935L..11L} and GRB\,230205A \citep{2023GCN.33309....1S}. 
Therefore, similar to the X-ray and optical bands, the detection of a radio afterglow can often be used to motivate deeper host-galaxy searches in the absence of a coincident host galaxy in shallower imaging.

Finally, we explore how our radio observations of short GRBs have improved in the last two decades, and the outlook for future observatories. Since the launch of {\it Swift} in 2004 \citep{2004ApJ...611.1005G}, 71 short or merger-driven GRBs with X-ray detected afterglows have been observed at radio wavelengths, the majority ($\sim 80\%$) of which were observed with the VLA. Prior to the conclusion of the VLA upgrade in 2012 \citep{2011ApJ...739L...1P}, only 2 short GRBs were detected at radio wavelengths \citep{2005Natur.438..988B, 2006ApJ...650..261S}, despite 27 being observed ($\sim 7\%$ detection rate). However, since the conclusion of the VLA upgrade, 15 of the 49 short/merger-driven  GRBs that have been observed at radio wavelengths have were detected \citep[][this work]{2014ApJ...780..118F, 2015ApJ...815..102F, 2017GCN.21395....1F, 2019ApJ...883...48L, 2021ApJ...906..127F, 2022ApJ...935L..11L, 2023GCN.33372....1S, 2023GCN.33358....1S, 
2023arXiv230810936S, 2023GCN.35097....1R}, resulting in a $\sim 34\%$ detection rate. Of those 15 radio-detected short GRBs, 13 were detected by the VLA. Thus, it is clear how the VLA's improved sensitivity has played an integral role in radio afterglow detection of short GRBs. Future planned radio facilities, such as the next-generation VLA \citep[ngVLA;][]{2015arXiv151006438C}, will provide observations an order of magnitude\footnote{\url{https://ngvla.nrao.edu/page/performance}} more sensitive than the current VLA. These facilities will not only allow us to detect the faintest short GRB afterglows, but also RS emission \citep[e.g.,][]{2017arXiv170908512L, 2018Galax...6..103L}.

\section{Conclusions}
\label{sec:Conclusions_231117A}

We have presented observations of the afterglow of \grb spanning 9 orders of magnitude in frequency using {\it Swift}, {\it CXO}, Keck, SOAR, T80-South, MMT,  ALMA, VLA, MeerKAT, and uGMRT. With detections extending to rest-frame times of 37 days, it is the longest-lived X-ray afterglow of a short GRB detected to date. Additionally, \grb is only the second short GRB to be detected with ALMA and MeerKAT, and the first to be detected by both. 
Motivated by the radio-afterglow detection of \grb, as well as the growing sample of radio-detected short GRBs, we explore the properties of these radio-bright bursts compared to the short GRB population as a whole.
We come to the following conclusions.
\begin{itemize}
    \item \grb is located at a small projected offset ($\sim 2$~kpc) from a star-forming host galaxy at $z = 0.257$, with stellar population properties that are typical for short GRB host galaxies.
    \item A standard FS can describe the broadband afterglow of \grb at $\gtrsim 1~$day. However, at $\lesssim 1~$day a refreshed FS with an RS is required. While we present two refreshed FS $+$ RS scenarios to explain the observed emission, it is difficult to account for the entire X-ray, optical, mm, and radio afterglows with either model.
    \item We find that radio-detected short GRBs are found at lower redshifts ($z<0.6$) with moderately brighter X-ray fluxes at $0.1~$days compared to the overall short GRB population. However, the X-ray luminosities of radio-detected short GRBs are similar to the short GRB population as a whole.
    \item Consistent with expectations from synchrotron afterglow theory, we find that the radio-detected short GRB population tends to be found at higher circumburst density environments ($> 10^{-2}~{\rm cm}^{-2}$), which is commensurate with their lower projected offsets ($< 8~$kpc), than the typical short GRB population. The afterglow properties of \grb are consistent with the radio-detected short GRB population.
    \item Radio detections of short GRBs are crucial for the determination of their jet collimation, especially in the cases of wide ($> 10^\circ$) jets, where the afterglow may only be detectable at radio wavelengths when the effects of collimation manifest in the light curves.
    \item In order to increase the likelihood of radio detection, we suggest that the depth of radio-afterglow searches should be increased for short GRBs with low X-ray fluxes ($< 10^{-12}{\rm erg~s}^{-1}{\rm cm}^{-2}$).
    \item Given the tendency for radio-detected short GRBs to occur in high-density environments close to their host's nucleus, we encourage deep observations for underlying host galaxies in the cases where a radio-bright short GRB appears highly offset from a candidate host galaxy.
\end{itemize}

Our work demonstrates the power of radio and mm observations spanning $\sim 1$--100~GHz in  studies of short GRB afterglows, and the impact that new or upgraded facilities (e.g., MeerKAT, ALMA, upgraded VLA) have had on their detection rate and the revelation  of additional emission components. 
Future radio facilities, such as the ngVLA, should be able to detect most short GRB afterglows, as well as capture RS emission. This will lead to a more robust understanding of these explosions and the environments in which they occur, as well as better constrain the collimation of short GRB jets, providing more accurate event rates. 

% \newpage
\section{Acknowledgements}

The authors thank Griffin Hosseinzadeh, Cesar Brice$\Tilde{\rm n}$o, Alexander Beckett, Wanjia Hu, and WeiKang Zheng for their contribution in acquiring the optical observations.
The Fong Group at Northwestern University acknowledges support by the U.S. National Science Foundation (NSF) under grants  AST-1909358, AST-2206494, AST-2308182, and CAREER grant  AST-2047919.
W.F. is grateful for support from the David and Lucile Packard Foundation, the Alfred P. Sloan Foundation, and the Research Corporation for Science Advancement through Cottrell Scholar Award \#28284.
W.J.-G.\ is supported by the NSF Graduate Research Fellowship Program under grant DGE-1842165. He also acknowledges support through National Aeronautics Space Administration (NASA) grants for {\it Hubble Space Telescope} programs GO-16075 and GO-16500.
Y.D. is supported by the NSF Graduate Research Fellowship Program under grant DGE-1842165.
T.E. is supported by NASA through the NASA Hubble Fellowship grant HST-HF2-51504.001-A awarded by the Space Telescope Science Institute, which is operated by the Association of Universities for Research in Astronomy, Inc., for NASA, under contract NAS5-26555.
Time-domain research by D.J.S.\ is supported by NSF grants AST-1908972, 2108032, and 2308181, and by the Heising-Simons Foundation under grant 2020-1864. 
A.R.E is supported by the European Space Agency Research Fellowship Program.
Support for this work was provided by NASA through Chandra award G03-24032X issued by the Chandra X-ray Center, which is operated by the Smithsonian Astrophysical Observatory for and on behalf of NASA under contract NAS8-03060.
A.V.F. is grateful for financial assistance from the Christopher R. Redlich Fund and numerous other donors.

Some of the data presented herein were obtained at the W. M. Keck Observatory, which is operated as a scientific partnership among the California Institute of Technology, the University of California, and NASA. The Observatory was made possible by the generous financial support of the W. M. Keck Foundation. The authors wish to recognize and acknowledge the very significant cultural role and reverence that the summit of Maunakea has always had within the indigenous Hawaiian community. We are most fortunate to have the opportunity to conduct observations from this mountain.
W. M. Keck Observatory and MMT Observatory access was supported by Northwestern University and the Center for Interdisciplinary Exploration and Research in Astrophysics (CIERA). Observations reported here were obtained at the MMT Observatory, a joint facility of the University of Arizona and the Smithsonian Institution.
Based in part on observations obtained at the Southern Astrophysical Research (SOAR) telescope, which is a joint project of the Minist\'{e}rio da Ci\^{e}ncia, Tecnologia e Inova\c{c}\~{o}es (MCTI/LNA) do Brasil,  NSF's NOIRLab, the University of North Carolina at Chapel Hill (UNC), and Michigan State University (MSU).
We thank Sean Points and the SOAR staff for coordinating our SOAR/Goodman observations.
The National Radio Astronomy Observatory is a facility of the NSF operated under cooperative agreement by Associated Universities, Inc.
This research was supported in part through the computational resources and staff contributions provided for the Quest high performance computing facility at Northwestern University which is jointly supported by the Office of the Provost, the Office for Research, and Northwestern University Information Technology.
Based in part on observations obtained through the Astronomical Event Observatory Network (AEON), a joint endeavor of the Las Cumbres Observatory and of NSF's NOIRLab, which is managed by the Association of Universities for Research in Astronomy (AURA) under a cooperative agreement with the U.S. NSF.
This work made use of data supplied by the UK Swift Science Data Centre at the University of Leicester.
We thank the staffs of all observatories mentioned above for their assistance in obtaining data.
The scientific results reported in this article are based in part on observations made by the Chandra X-ray Observatory. This research has made use of software provided by the Chandra X-ray Center (CXC) in the application package CIAO. The MeerKAT telescope is operated by the South African Radio Astronomy Observatory, which is a facility of the National Research Foundation, an agency of the Department of Science and Innovation. We thank the staff of the GMRT that
made these observations possible. GMRT is run by the
National Centre for Radio Astrophysics of the Tata Institute
of Fundamental Research.
This paper makes use of the following ALMA data: ADS/ JAO.ALMA\#2022.1.00624.T. ALMA is a partnership of ESO (representing its member states), NSF (USA) and NINS (Japan), together with NRC (Canada), MOST and ASIAA (Taiwan), and KASI (Republic of Korea), in cooperation with the Republic of Chile. The Joint ALMA Observatory is operated by ESO, AUI/NRAO and NAOJ. 

\newpage

\facilities{VLA, MeerKAT, GMRT, Swift, CXO, Keck:I (LRIS), MMT (MMIRS), SOAR (Goodman), T80-South Telescope}

\software{
{\tt CASA} \citep{2007ASPC..376..127M}, 
{\tt CIAO} \citep{2006SPIE.6270E..1VF}, 
{\tt DoPhot} \citep{1993PASP..105.1342S},
{\tt dynesty} \citep{2020MNRAS.493.3132S},
{\tt emcee} \citep{2013PASP..125..306F},
{\tt hotpants} \citep{2015ascl.soft04004B},
{\tt matplotlib} \citep{2007CSE.....9...90H},
{\tt pandas} \citep{pandas}, 
{\tt Prospector} \citep{2017ApJ...837..170L}, 
{\tt pwkit} \citep{2017ascl.soft04001W}, 
{\tt scipy} \citep{scipy},
{\tt sextractor} \citep{sextractor},
{\tt swarp} \citep{2010ascl.soft10068B}
}

\newpage 
\bibliographystyle{apj}
\bibliography{library,journals_apj}

\clearpage

\end{document}